\documentclass{cjps}
\usepackage{fancyhdr}
\usepackage{graphicx}
\usepackage[colorlinks,bookmarks,%dvipdfm,
urlcolor=black,linkcolor=black,anchorcolor=black,citecolor=black]{hyperref}  %%%%fhh dvipdfm will cause some problems

\newcommand{\page}{1}% starting page number
\newcommand{\receivedate}{October 14, 2015}%%%jmn Here the date was Decemeber 26, 2014, it was corrected.
\newcommand{\thisyear}{2015}
\newcommand{\thismonth}{November}
\newcommand{\thetitle}{General Poincar\'e Gauge Theory Cosmology}
\newcommand{\theshorttitle}{\uppercase{General Poincar\'e Gauge Theory Cosmology
 }%\ldots
 }
\newcommand{\theauthor}{\uppercase{F.-H. Ho, H. Chen, J. M. Nester, and H.-J. Yo}}
\newcommand{\refdoi}[1]{doi: \href{http://dx.doi.org/#1}{#1}}
\newcommand{\thems}{\thepage}

\fancypagestyle{cjp}{ \fancyhf{}
\fancyhead[L]{\small\textrm{}}%{CHINESE JOURNAL OF PHYSICS}} %%%%fhh comment for arXiv
\fancyhead[C]{\small\textrm{}}%{VOL.~\volume, NO.~\issue}}%%%%fhh comment for arXiv
\fancyhead[R]{\small\textrm{\thismonth{} \thisyear{}}}
\fancyfoot[L]{\small\textrm{}}%{http://PSROC.phys.ntu.edu.tw/cjp}} %%%%fhh for arXiv
\fancyfoot[C]{\small\textbf\thems}
\fancyfoot[R]{\small\textrm{}}%{
}

\begin{document}
\setcounter{page}{\page}
\title{\thetitle}
\received{\receivedate}
%\revised{\revisedate}  %%%jmn Removed.

\author{Fei-Hung Ho}
%\email{fei@ntnu.edu.tw}%%%jmn Old email dropped
\email{fhho@zjut.edu.cn}%%%jmn Updated email, as requested by author.
\affiliation{Institute for Advanced Physics and Mathematics, Zhejiang University of Technology, Hangzhou, China}

\author{Hsin Chen}
\email{eppp8ster@gmail.com}
\affiliation{Department of Mathematics and Science, College of International Studies and Education for Overseas Chinese Students, National Taiwan Normal University, Taipei, Taiwan}

\author{James M. Nester}
\email{nester@phy.ncu.edu.tw}
\affiliation{Department of Physics \& Center for Mathematics and Theoretical Physics, National Central University, Chungli 320, Taiwan}
\affiliation{Graduate Institute of Astronomy, National Central University, Chungli 320, Taiwan}
\affiliation{Leung Center for Cosmology and Particle Astrophysics, National Taiwan University, Taipei 10617, Taiwan}
\affiliation{Morningside Center of Mathematics, Academy of Mathematics and System Sciences, Chinese Academy of Sciences, Beijing, 100190, China}

\author{Hwei-Jang Yo}
\email{hjyo@phys.ncku.edu.tw}
\affiliation{Department of Physics, National Cheng-Kung University, Tainan City, Taiwan, Republic of China}

\begin{abstract}
For the quadratic Poincar\'e gauge theory of gravity (PG) we consider the FLRW cosmologies using an isotropic Bianchi representation.  Here the considered cosmologies are for the general case: all the even and odd parity terms of the quadratic PG with their respective scalar and pseudoscalar parameters are allowed with no \emph{a priori} restrictions on their values.  With the aid of a manifestly homogeneous and isotropic representation, an effective Lagrangian gives the second order dynamical equations for the gauge potentials.  An equivalent set of first order equations for the observables is presented.  The generic behavior of physical solutions is discussed and illustrated using numerical simulations.
\end{abstract}

%\doi{10.6122/CJP.20151014}

\maketitle
\thispagestyle{cjp}

\section{Introduction}

All the known fundamental physical interactions can be formulated
in a common framework: as {\em local gauge theories}.
This is true especially of gravity; however Einstein's general relativity (GR), when
 viewed as a theory for a dynamic spacetime metric, does not so well reveal its affinity with other gauge theories. (For a recent discussion of the fundamental gauge theory nature of gravity, see Ref.~\cite{GR100}.)
Physically (and geometrically) it is reasonable to consider gravity as a
gauge theory of the local Poincar\'e symmetry of Minkowski
spacetime.  A formulation of gravity based on local spacetime geometry gauge symmetry, the quadratic Poincar\'e gauge theory of gravity (PG, a.k.a.~PGT) was worked out some time ago~\cite{HHKN,Hehl80,HS80,MieE87,HHMN95,GFHF96,Blag02}, and now there is a comprehensive reader with reprints and comments~\cite{BlagoHehl}.  The PG is formulated in terms of Riemann-Cartan spacetime geometry, having a metric and a metric compatible connection.  Such a connection has, in general, both curvature and torsion.

%\subsection{The search for good dynamic PG modes}

Independent of the metric, the connection has six possible dynamic connection modes; they carry spin-$2^{\pm}$, spin-$1^{\pm}$, spin-$0^{\pm}$.  A good dynamic
mode should transport positive energy and should not propagate
outside the forward null cone.  Investigations (especially Refs.~\cite{HS80,SN80}) of the linearized (even parity) quadratic
PG theory found
that at most three modes can be simultaneously dynamic; all the
acceptable cases were tabulated; many combinations of three modes
are satisfactory to linear order. Complementing this, the
Hamiltonian analysis revealed the related constraints~\cite{BMNI83}.
Then detailed investigations of the Hamiltonian and propagation~\cite{CNY98,HNZ96,yo-nester-99,yo-nester-02} concluded that effects
due to nonlinearities in the constraints could be expected to render
all of these cases physically unacceptable except for the two
``scalar modes'', carrying spin-$0^+$ and spin-$0^-$.

 In order to further investigate the dynamical possibilities of these PG scalar modes, Friedmann-Lema\^{i}tre-Robertson-Walker (FLRW)
 cosmological models were considered.  Using a $k=0$ model it was found that the $0^+$ mode naturally couples to the acceleration of the universe and could account for the present day observations~\cite{YN07,SNY08}; this model was then extended to include the $0^-$ mode~\cite{JCAP09}.

There is no known fundamental reason why the gravitational coupling should respect parity.
Odd parity terms with pseudoscalar parameters were first introduced into the quadratic PG some time ago~\cite{OPZ}, but this innovation was not followed up until more recent times.
Now there is renewed interest in all the possible couplings between even and odd parity modes. The odd parity terms with their pseudoscalar coupling parameters were rediscovered and included in the PG~\cite{BHN,diakonov,BH}.
After systematically developing the general parity PG theory, in BHN~\cite{BHN} the two scalar torsion mode PG Lagrangian was extended to include the appropriate pseudoscalar coupling constants that provide cross parity coupling (often referred to as ``parity violating'' terms) and the FLRW cosmological models were formulated.  The BHN cosmologies were then further explored~\cite{iard10,HN,AdP}.

The dynamics of the PG BHN model was expected to be clearly revealed in purely time dependent solutions, hence {\em homogeneous cosmologies} were investigated.
The two dynamical connection modes carry spin 0$^+$ and spin 0$^-$ (referred to as {\em scalar} modes or, more specifically, as the {\em scalar} and {\em pseudoscalar} mode).
Consequently in a homogeneous situation they cannot pick out any spatial direction,
and thus they have no interaction with spatial anisotropy,
so for a study of their dynamics it is most simple and appropriate to look to {\em isotropic} models.
 For the PG BHN model, following the technique used in~\cite{JCAP09},
 an effective Lagrangian and Hamiltonian as well as a system of first order dynamical equations
 for Bianchi class A isotropic homogeneous cosmological models was constructed,
 and some sample evolution was presented which showed the effect of the cross parity coupling.
The normal modes were identified, and it was shown analytically how they control the late time asymptotics.  A numerical evolution example was presented which shows that the asymptotic late time normal mode evolution is a good approximation~\cite{iard10,HN,AdP}.

Recently Karananas took on the major task of analyzing the modes in the general quadratic PG including all the odd parity terms in the Lagrangian.  At first it seemed that the general PG theory (including all the scalar and pseudoscalar parameters) would allow for all the possible connection modes to have \emph{good} (i.e., ``no ghosts, no tachyons'') propagation.  However, more careful checking showed that this is not possible after all~\cite{Karananas}, nevertheless it does seem that the healthy parameter space of the theory is extended by the inclusion of the odd parity terms. In view of this it is timely to reconsider cosmology for the general quadratic PG, with no \emph{a priori} restrictions on the parameters.

We would like to note that the PG theory is a viable alternative to GR.  While there are no observations that confirm non-vanishing torsion, there are no observations that exclude torsion, only some mild constraints on the magnitude.
For a detailed discussion and a recent assessment of the observational status of torsion the reader may consult Ni~\cite{ni}.  Briefly, we just mention that torsion naturally couples to spin, to directly detect it may require highly spin polarized materials.  On the other hand it may be that torsion could play a significant role on the cosmological scale.

%\subsection{Torsion Cosmology}
Since the 1970s cosmological
models for the Einstein-Cartan (EC) theory
and the Poincar\'e gauge theory of gravity (PG) have been developed. For the EC theory see, e.g., Kopczyn\'ski~\cite{Kopczynski}, Trautman~\cite{Trautman},
Tafel~\cite{Tafel}, and Kuchowicz~\cite{Kuch}.  The special interest then was
the possibility that models with torsion could avoid singularities, however it was soon noted that torsion is likely to make singularities more severe rather than prevent them, see, e.g., Kerlick~\cite{Kerlick} and~\cite{NesIsen77}.
Minkevich \textit{et al}.~\cite{Min8083,MN95,MG06,MGK07} developed
 torsion cosmology in a series of papers. For the early work on PG cosmology see~Goenner and M\"uller-Hoissen~\cite{GMH}.  A more recent report on the status
of the subject was given by Puetzfeld~\cite{Peutzfeld}, which surveyed virtually all of the papers on the subject up to 2004; as far as we know it only overlooked~\cite{NesIsen77}.
For the recent work by our group see~\cite{YN07,SNY08,JCAP09,iard10,HN,AdP,ChenThes}.
For other recent works see~\cite{WW09,Pop10,ALX10,AL12,LSX09,GLT,TLG}.

%\subsection{Overview}
The objective of this present work is to begin the analysis of the \emph{general quadratic PG cosmology}.
Specifically we consider homogeneous isotropic cosmologies, using an isotropic Bianchi representation to obtain a manifestly homogeneous formulation.  This allows us to obtain the dynamical equations via an effective Lagrangian.  We find a set of 6 first order equations and discuss the behavior of generic solutions showing some typical numerical evolution.

\section{the Poincar\'e Gauge Theory}

\subsection{Riemann-Cartan geometry}

The Poincar\'e gauge theory of gravity (PG)~\cite{HHKN,Hehl80,HS80,MieE87,HHMN95,GFHF96,Blag02,BlagoHehl} is based on spacetime with a Riemann-Cartan geometry, i.e., a Lorentz signature metric with a metric compatible connection.  The two sets of gauge potentials are, respectively, for \emph{translations} the orthonormal coframe and for \emph{Lorentz-rotations} the metric compatible (Lorentz Lie algebra valued) connection one-forms:
\begin{equation}\vartheta^\alpha=e^\alpha{}_i dx^i,\qquad
\Gamma^{\alpha\beta}=\Gamma^{[\alpha\beta]}{}_idx^i.
\end{equation}
The associated field strengths are the \emph{torsion} and \emph{curvature} 2-forms:
\begin{eqnarray}
T^\alpha&:=&d\vartheta^\alpha+\Gamma^\alpha{}_\beta\wedge\vartheta^\beta=\frac12 T^\alpha{}_{\mu\nu} \vartheta^\mu\wedge \vartheta^\nu,\label{torsion2}\\
R^\alpha{}_\beta&:=&d\Gamma^\alpha{}_\beta+\Gamma^\alpha{}_\gamma\wedge\Gamma^\gamma{}_\beta=\frac12 R^\alpha{}_{\beta\mu\nu}\vartheta^\mu\wedge\vartheta^\nu,\label{curv2}
\end{eqnarray}
which satisfy the respective first and second \emph{Bianchi identities}:
\begin{equation}
DT^\alpha\equiv R^\alpha{}_\beta\wedge\vartheta^\beta, \qquad DR^\alpha{}_\beta\equiv0.\label{BianchiId}
\end{equation}
The metric is
$g=-\vartheta^0\otimes\vartheta^0+\delta_{ab}\vartheta^a\otimes\vartheta^b$, $a,b=1,2,3$.
The coframe along with the Hodge dual gives a convenient dual basis for forms: $\eta^{\alpha\beta\dots}:=*\vartheta^{\alpha\beta\dots}$, where $\vartheta^{\alpha\beta\dots}:=\vartheta^\alpha\wedge\vartheta^\beta\cdots$.
The volume 4-form is $\eta:=*1=(4!)^{-1}\eta_{\alpha\beta\mu\nu}\vartheta^{\alpha\beta\mu\nu}\equiv\vartheta^{0123}$, with $\eta_{\alpha\beta\mu\nu}$ being the totally anti-symmetric Levi-Civita tensor.

\subsection{General Lagrangian}

The Lagrangian density for the PG is taken to have a Yang-Mills type form, including up to quadratic terms in the field strengths. It has the
general structure
\begin{eqnarray}\label{quadraticL}
{\cal L}_{\rm PG} \sim \frac{1}{\kappa}\left(\Lambda+ \text{curvature}
  +\text{torsion}^2\right) + \frac{1}{\varrho}\,\text{curvature}^2\,,
\end{eqnarray}
which includes two types of terms with different physical dimensions:
 $\kappa:=8\pi G/c^4$ is the usual gravitational constant, and $\varrho^{-1}$ has the dimensions of action. $\Lambda$ is the cosmological constant.
The associated field equations for the coframe and connection are obtained by varying with respect to these potentials.  This gives dynamical equations for the potentials of the quasi-linear second order type
 with the respective material sources (obtained from some ${\cal L}_\psi={\cal L}(\vartheta^\alpha,\psi,D\psi)$, a minimally coupled matter Lagrangian) being the energy-momentum, $\sim\delta{\cal L}_\psi/\delta\vartheta^\mu$, and the spin density $\sim\delta{\cal L}_\psi/\delta\Gamma^{\alpha\beta}$ 3-forms.  The dynamical equations have the general form
\begin{eqnarray}
\kappa^{-1}(\Lambda+ \hbox{curvature} + D \hbox{ torsion} + \hbox{torsion}^2)+ \varrho^{-1}\hbox{curvature}^2&=& \hbox{energy-momentum},\qquad\\
\kappa^{-1}\hbox{torsion}+\varrho^{-1} D\hbox{ curvature}&=& \hbox{spin}.
\end{eqnarray}
From these two equations, with the aid of the Bianchi identities~(\ref{BianchiId}) (and also via a Noether theorem type argument) one can obtain the conservation of energy-momentum and angular momentum expressions.

In complete detail the general quadratic PG Lagrangian 4-form is~\cite{BHN}
\begin{eqnarray}\nonumber
{\cal L}_{\rm PG} &=&
  \frac{1}{2\kappa}\Bigl(a_0 R\eta+b_0 X\eta
    -2\Lambda\eta +
     \textstyle\sum
    \limits_{I=1}^{3}a_{I}{}^{(I)}T^\alpha\wedge *{}^{(I)}
    T_\alpha\Bigr)\nonumber\\
  &&  + \frac{1}{{\kappa}}\left( {\sigma}_{1}{}
^{(1)}T^{\alpha}\wedge\, ^{(1)}T_{\alpha} + {\sigma}_{2}{}
^{(2)}T^{\alpha}\wedge{} ^{(3)}T_{\alpha}\right)\cr
     & & -\frac{1}{2\varrho} %R^{\alpha\beta}\wedge
  \Bigl(\textstyle\sum\limits_{I=1}^{6}w_{I}{}^{(I)}R^{\alpha\beta}\wedge*{}^{(I)}
  R_{\alpha\beta}\Bigr)\,\nonumber \\ %;
&&-\frac{1}{2{\varrho}}\left(\ {\mu}_{1}{} ^{(1)}R^{\alpha\beta}\wedge{}
^{(1)}R_{\alpha\beta} + {\mu}_{2}{} ^{(2)}R^{\alpha\beta}\wedge{}
^{(4)}R_{\alpha\beta} \right. \nonumber\\ & & %\hspace{-7pt}
\qquad + \left.
{\mu}_{3}{}^{(3)}R^{\alpha\beta}\wedge{} ^{(6)}R_{\alpha\beta} +
{\mu}_{4}{} ^{(5)}R^{\alpha\beta}\wedge{} ^{(5)}R_{\alpha\beta}
\right).\label{PGgeneralL}
\end{eqnarray}
Here $R$ is the scalar curvature and $X$ is the pseudoscalar curvature ($X\equiv -\frac12R_{\alpha\beta\mu\nu}\eta^{\alpha\beta\mu\nu}$).  The torsion has been decomposed into three algebraically irreducible pieces: $T^\alpha={}^{(1)}T^\alpha+{}^{(2)}T^\alpha+{}^{(3)}T^\alpha$, which are, respectively, a pure tensor (16 components), the trace (vector), and a totally antisymmetric part (dual to an axial vector). Similarly the curvature 2-form has been decomposed into a sum of 6 algebraically irreducible pieces: $R^\alpha{}_\beta=\sum_{I=1}^6{}^{(I)}R^\alpha{}_\beta$, namely, in numerical order: weyl, pair-commutator, pseudoscalar, ricci-symmetric, ricci-antisymmetric, and scalar.  The respective number of components is (10,9,1,9,6,1).
In the above Lagrangian the parameters $\Lambda,a_0,a_I,w_I$ (which multiply even parity 4-forms) are scalars, and the parameters $b_0,\sigma_I,\mu_I$ (which multiply odd parity 4-forms) are pseudoscalars.
The general theory has 11 scalar plus 7 pseudoscalar parameters.  But they are not all physically independent.
They are subject to 1 even parity and 2 odd total differentials, leaving effectively 10 scalar + 5 pseudoscalar = 15 ``physical'' parameters, as we will briefly explain, referring to Refs.~\cite{BHN,BH} for details.

\subsection{Topological terms}

Not all of the above parameters are physically independent, since there are 3 topological invariants.
Without changing the field equations, one can add to the Lagrangian 4-form~(\ref{PGgeneralL}) any multiple of the (odd parity) \emph{Nieh-Yan identity}~\cite{NY}:
\begin{equation}
 T^\alpha\wedge T_\alpha-R_{\alpha\beta}\wedge\vartheta^{\alpha\beta} \equiv  d(\vartheta^\alpha\wedge T_\alpha).
 \end{equation}
 From the irreducible decomposition given in Eqs.~(6,7) in~\cite{BH}, one can see that this allows one to make parameter changes of the type
\begin{equation}
\Delta(b_0,\sigma_1,\sigma_2)=(-f_1,f_1/2,f_1).\label{nieh-yan}
\end{equation}
Also one can add a multiple of the (even parity) \emph{Euler 4-form}  $R^{\alpha\beta}\wedge R^{\gamma\delta}\eta_{\alpha\beta\gamma\delta}$.  Because of the 2nd Bianchi identity~(\ref{BianchiId}b), this makes no contributions to the field equations.  From the irreducible decomposition of Eqs.~(14b,16,18) in~\cite{BH}, it can be seen that this induces parameter changes of the type
\begin{equation}
\Delta(w_1,w_2,w_3,w_4,w_5,w_6)=(-f_3,f_3,-f_3,f_3,-f_3,-f_3).\label{euler}
\end{equation}
Furthermore once can add a multiple of the (odd parity) \emph{Pontryagin} 4-form $R^\alpha{}_\beta\wedge R^\beta{}_\alpha$. Again, thanks to the 2nd Bianchi identity, this has no effect on the field equations. For the irreducible decomposition of Eqs.~(14a,16,20) in~\cite{BH}, one sees that this induces parameter changes of the type
\begin{equation}
\Delta(\mu_1,\mu_2,\mu_3,\mu_4)=(-f_2,-2f_2,-2f_2,-f_2).\label{pointryagin}
\end{equation}
The actual physical equations will only depend on combinations of the 18 parameters that are invariant under such transformations.

\subsection{Minimal versions}

The topological identities could be used to eliminate some parameters and thus simplify computations.
Baekler and Hehl used the Nieh-Yan, Euler, and Pontryagin forms to respectively eliminate the parameters $\sigma_1,\ w_1,\ \mu_1$ and arrive at their Lagrangian (see Eq.~(56) in~\cite{BH}); it has no explicit weyl curvature terms, ${}^{(1)}R^\alpha{}_\beta$ (this is the largest curvature part, having 10 components).
 One of the many alternatives is to instead use the Euler and Pontryagin forms to respectively eliminate the parameters $w_2,\ \mu_2$; then the resulting Lagrangian would not contain any terms involving ``paircom'' ${}^{(2)}R^\alpha{}_\beta$ (a less familiar part of the curvature with 9 components).
  The decision is simply a practical question, not a fundamental issue: one could select the most convenient choice for a particular application, although this complicates the comparison of results obtained via different choices.

The computations presented here are completely general: all the PG parameters are allowed to have any values, no special restrictions or choices have been imposed.

\section{Generic PG cosmology kinematics}
Following~\cite{HN,AdP,ChenThes},
for homogeneous, isotropic Bianchi type I and IX
(respectively equivalent to FLRW $k=0$ and $k=+1$) cosmological models we take the
isotropic orthonormal coframe to have the form
\begin{equation}
\vartheta^0:=dt,\qquad \vartheta^a:=a\sigma^a.
\end{equation}
Here $a=a(t)$ is the scale factor and $\sigma^a$ depends on the (not needed here) spatial coordinates in such a way that
\begin{equation}
d\sigma^a=\zeta\epsilon^a{}_{bc}\sigma^b \wedge\sigma^c
\end{equation}
(here $\epsilon_{abc}\equiv\epsilon_{[abc]}$ is the 3D Levi-Civita symbol),
where $\zeta=0$ for Bianchi I and $\zeta=1$ for Bianchi IX, thus $\zeta^2=k$,
the FLRW spatial Riemannian curvature parameter.
(We remark that our \emph{derivation} here of the dynamical cosmological equations applies only to the cases $k=0$ and 1.
The $k=-1$ case does not admit a manifestly isotropic representation and needs to be treated separately~\cite{ChenThes,MT72}.  In fact, the dynamical equations obtained here are actually valid also for $k=-1$; the details justifying this will be presented elsewhere.)

As a consequence of isotropy, the only non-vanishing connection one-form
coefficients are necessarily of the form
\begin{equation} \Gamma^a{}_0=\psi(t)\,
\sigma^a,\qquad \Gamma^a{}_b=\chi(t)\epsilon^a{}_{bc}\, \sigma^c.\end{equation}

From the definition~(\ref{curv2}), $R^\alpha{}_\beta:=d\Gamma^\alpha{}_\beta+\Gamma^\alpha{}_\gamma\wedge\Gamma^\gamma{}_\beta$,
 all the independent nonvanishing
curvature 2-form components are found to be
\begin{eqnarray}
R^{ab}&=&\dot \chi dt\wedge
\epsilon^{ab}{}_{c}\sigma^c+[\psi^2-(\chi-\zeta)^2+\zeta^2]\sigma^a\wedge \sigma^b,\\
R^a{}_0&=&\dot \psi dt\wedge \sigma^a
-\psi(\chi-\zeta) \epsilon^a{}_{bc}\sigma^b\wedge\sigma^c
.
\end{eqnarray}
Consequently, the scalar and pseudoscalar curvatures are,
respectively,
\begin{eqnarray}
R&=&6[a^{-1}\dot \psi+a^{-2}(\psi^2-[\chi-\zeta]^2+\zeta^2)], \label{R} \\
X&=&6[a^{-1}\dot \chi+2a^{-2}\psi(\chi-\zeta)].\label{X}
\end{eqnarray}

Regarding the other irreducible parts of the curvature, for FLRW cosmogies we know that parts 1 and 5 (Weyl and antisymmetric Ricci) identically vanish~\cite{BHN}.
For part 4 (traceless Ricci symmetric) we find
\begin{equation}
^{(4)}R^{0b}{}_{0c}=\frac1{12} \tilde R\delta^b_c,\qquad
^{(4)}R^{ab}{}_{cd}=-\frac1{12} \tilde R\delta^{ab}_{cd},
\end{equation}
\begin{equation}
\tilde R:=6[a^{-1}\dot\psi-a^{-2}(\psi^2-[\chi-\zeta]^2+\zeta^2)].
\end{equation}
For part 2 (paircom) we find
\begin{eqnarray}
^{(2)}R_{ab0c}&=&-{}^{(2)}R_{0cab}=\frac1{12} \tilde X\epsilon_{abc},\\
\tilde X&:=&6[a^{-1}\dot\chi-2a^{-2}\psi(\chi-\zeta)].
\end{eqnarray}
 Via the torsion expressions that will be given next, these results are equivalent to  Eqs.~(147) and (149) in BHN~\cite{BHN}.

Because of isotropy, the only nonvanishing torsion tensor
components are of the form
\begin{equation}
T^a{}_{0b}=u(t)\delta^a_b, \qquad T^a{}_{bc}=-
2x(t)\epsilon^a{}_{bc},
\end{equation}
where $u$ and $x$ are referred to as the {\em scalar\/} and {\em pseudoscalar\/} torsion, respectively.
They occur in the vector and axi-vector parts of the torsion.  The tensor part, ${}^{(1)}T^\alpha$, vanishes for isotropic cosmologies, so it will play no role in our considerations.  From the torsion 2-form definition~(\ref{torsion2}), $T^\mu:=D\vartheta^\mu:=d\vartheta^\mu+\Gamma^\mu{}_\nu\wedge\vartheta^\nu$, the
relations between these torsion components and the gauge variables are found to be
\begin{equation}
u=a^{-1}(\dot a-\psi), \qquad x= a^{-1}(\chi-\zeta). \label{fchi}
\end{equation}
In terms of the \emph{Hubble function}, $H:=a^{-1}\dot a$, we have
\begin{equation}a^{-1}\psi=H-u,\label{Hu}\end{equation}
 a relation which clearly shows that {\em kinematically\/} the scalar torsion $u$ {\em directly couples\/} to the rate of {\em expansion\/} of the universe.

 From the symmetry assumptions of our model, the source material energy-momentum tensor is {\em necessarily\/} of the fluid form. Thus it can be described by an effective energy density $\rho$, pressure $p$, and flow vector along the cosmological time axis.  Here we assume that the source spin density is negligible; this is a quite reasonable physical assumption except for the very early universe.

\subsection{Quadratic curvature terms}

For isotropic cosmology, since the curvature pieces 1 and 5 vanish,
 the quadratic curvature terms in the general Lagrangian~(\ref{PGgeneralL}) involving these pieces (i.e., those with the parameters $w_1,\ w_5,\ \mu_1,\ \mu_4$) are not relevant.

The BHN model investigations~\cite{BHN,HN,AdP} considered only those quadratic curvature terms involving the scalar and pseudoscalar curvature, $R$ and $X$.
The quadratic terms involving these pieces (with the parameters $w_3,\ w_6,\ \mu_3$) were the only ones included in the BHN Lagrangian.
In comparison with the BHN model considered in those earlier studies the general quadratic PG (gqPG) cosmological Lagrangian 4-form gains three new quadratic curvature terms:
\begin{equation}
-\frac{1}{2\varrho}\left[w_2 {}^{(2)}R_{\alpha\beta} \wedge * {}^{(2)}R^{\alpha\beta}
+w_4 {}^{(4)}R_{\alpha\beta} \wedge * {}^{(4)}R^{\alpha\beta}
+\mu_2 {}^{(2)}R_{\alpha\beta} \wedge {}^{(4)}R^{\alpha\beta}
\right].
\end{equation}
For homogeneous-isotropic cosmology the respective values are
\begin{eqnarray}
 {}^{(2)}R_{\alpha\beta} \wedge * {}^{(2)}R^{\alpha\beta}&=&-\frac1{12}\tilde X^2\eta,\\
 {}^{(4)}R_{\alpha\beta} \wedge * {}^{(4)}R^{\alpha\beta}&=&\frac1{12}\tilde R^2\eta,\\
 {}^{(2)}R_{\alpha\beta} \wedge {}^{(4)}R^{\alpha\beta}&=&-\frac1{12}\tilde X\tilde R\eta.
\end{eqnarray}
Here $\eta=a^3 dt\wedge \sigma^1\wedge\sigma^2\wedge\sigma^3$ is the proper 4-volume 4-form.

\subsection{Topological terms in cosmology}

Let us see in detail how certain combinations of terms add up to a total time derivative for these cosmologies.

For the even parity Euler quadratic curvature terms, the pattern from~(\ref{euler}) is of the form $|{}^{(2)}R|^2-|{}^{(3)}R|^2+|{}^{(4)}R|^2-|{}^{(6)}R|^2$:
\begin{eqnarray}
\frac{a^3}{12}\{-\tilde X^2+X^2+\tilde R^2-R^2\}&\equiv&3a^3\big\{-
[a^{-1}\dot\chi-2a^{-2}\psi(\chi-\zeta)]^2\nonumber\\
&&\quad+[a^{-1}\dot \chi+2a^{-2}\psi(\chi-\zeta)]^2\nonumber\\
&&\quad+[a^{-1}\dot\psi-a^{-2}(\psi^2-\chi^2+2\zeta\chi)]^2\nonumber\\
&&\quad-[a^{-1}\dot \psi+a^{-2}(\psi^2-\chi^2+2\zeta\chi)]^2\big\}\nonumber\\
&\equiv&12[2\dot\chi\psi(\chi-\zeta)-\dot\psi(\psi^2-\chi^2+2\zeta\chi)],
\end{eqnarray}
the quadratic parts cancel and the cross terms add up to an expression which is a total time derivative of $12\psi\chi^2-4\psi^3-24\zeta\psi\chi$.

For the odd parity Pontryagin form the pattern is $\langle{}^{(2)}R\times {}^{(4)}R\rangle+\langle{}^{(3)}R\times {}^{(6)}R\rangle$:
\begin{eqnarray}
\frac{a^3}{12}\{-\tilde X\tilde R+RX\}&\equiv&3a^3\big\{[-a^{-1}\dot\chi+2a^{-2}\psi(\chi-\zeta)]
[a^{-1}\dot\psi-a^{-2}(\psi^2-\chi^2+2\zeta\chi)]\nonumber\\
&&\quad+[a^{-1}\dot \psi+a^{-2}(\psi^2-\chi^2+2\zeta\chi)] [a^{-1}\dot\chi+2a^{-2}\psi(\chi-\zeta)]\big\}\nonumber\\
&\equiv&6[\dot\chi(\psi^2-\chi^2+2\zeta\chi)+2\psi\dot\psi(\chi-\zeta)],
\end{eqnarray}
which again is a total time derivative.

\section{Effective Lagrangian}

For homogenous cosmologies for our models (which are in Bianchi class A), as in GR~\cite{MT72}, one can obtain an effective Lagrangian by integrating over space.
Our {\em effective Lagrangian} for the gqPG $L_{\mathrm{eff}}=L_\mathrm{PG}+L_{\mathrm{int}}$
 includes the {\em \textit{interaction}} Lagrangian:
$L_{\mathrm{int}}= pa^3$, where $p=p(t)$ is the source fluid pressure.
The gravitational Lagrangian
for the gqPG cosmology is
\begin{eqnarray}
L_\mathrm{PG} &=& \frac{1}{2\kappa}(a_0R+b_0X-2\Lambda)a^3+\frac3{2\kappa}(-a_2u^2+4a_3x^2-4\sigma_2ux)a^3\nonumber\\
               &&-\frac1{24\varrho}[-w_2\tilde X^2-w_3X^2+w_4\tilde R^2+w_6R^2-\mu_2\tilde X\tilde R+\mu_3RX]a^3.\quad \label{generalL}
\end{eqnarray}
(Remark: The sign of the $\sigma_2$ term in the gravitational Lagrangian here differs from that used in our earlier works~\cite{HN,AdP}.  This adjustment has been made
in order to conform with the convention established in~\cite{BHN}. Various formulas below correct some other sign errors in~\cite{HN,AdP} for which \emph{errata} are in preparation.)
Although our model makes sense for all values of the parameters, there are certain inequalities that identify what can be regarded as the \emph{physical} range.
For \emph{least action} one should have \emph{positive} ``kinetic energy'', hence the terms with quadratic time derivatives should have positive coefficients, which yields the requirements
\begin{equation}a_2<0,\quad w_2+w_3>0, \quad w_4+w_6<0, \quad 4 (w_2+w_3)(w_4+w_6)+(\mu_3-\mu_2)^2<0.\end{equation}

In the following, for simplicity, we often take units such that $\kappa=1=\varrho$.
  In the final results these factors can be easily restored by multiplying $\{a_0,a_2,a_3,b_0,\Lambda,\sigma_2\}$ by $\kappa^{-1}$ and
$\{w_2,w_3,w_4,w_6,\mu_2,\mu_3\}$ by $\varrho^{-1}$.

\subsection{Useful combinations}

 In the general effective Lagrangian~(\ref{generalL}),
 by using the fact that $\tilde R^2,\ \tilde X^2,\ \tilde X\tilde R$ differ from $R^2,\ X^2,\ XR$ only by having opposite sign cross terms,
 the
quadratic curvature terms can be re-expressed
as follows:
\begin{eqnarray}
&&-\frac1{24}[-w_2\tilde X^2-w_3X^2+w_4\tilde R^2+w_6R^2-\mu_2\tilde X\tilde R+\mu_3RX]a^3\nonumber\\
&&\qquad\equiv-\frac1{24}[-w_{3+2}X^2+w_{6+4}R^2+\mu_{3-2}RX]a^3 \nonumber\\
&&\qquad\qquad-\frac1{24}[-w_2(\tilde X^2-X^2)+w_4(\tilde R^2-R^2)-\mu_2(\tilde X\tilde R-XR)
]a^3\nonumber\\
&&\qquad\equiv -\frac1{24}[-w_{2+3}X^2+w_{4+6}R^2+\mu_{3-2}RX]a^3\nonumber\\
&&\qquad\qquad -12w_2\dot\chi\psi(\chi-\zeta)
 +6{w_4}\dot\psi[\psi^2-(\chi-\zeta)^2+\zeta^2]\nonumber\\
&&\qquad\qquad -3\mu_2[2\psi(\chi-\zeta)\dot\psi+(\psi^2-\chi^2+2\chi\zeta)\dot\chi]\nonumber\\
&&\qquad \equiv-\frac1{24}[-w_{2+3}X^2+w_{4+6}R^2+\mu_{3-2}RX]a^3\nonumber\\
&&\qquad\qquad +3w_{4-2}[-\dot\psi(\chi-\zeta)^2+2\dot\chi\psi(\chi-\zeta)]%\nonumber\\
 +\hbox{total time derivative}.\label{quadcurv}
\end{eqnarray}
Here, in order to have a more compact presentation, we introduced certain convenient parameter combinations:
\begin{equation}
w_{2+3}:=w_2+w_3, \quad w_{4-2}:=w_4-w_2,\quad w_{4+6}:=w_4+w_6, \quad \mu_{3-2}:=\mu_3-\mu_2,
\end{equation}
which are invariant under the allowed topological transformations~(\ref{euler}), (\ref{pointryagin}) and hence are \emph{physically meaningful}.

 \subsection{Energy function}

The gqPG energy function can be readily obtained from the earlier work~\cite{HN} Eq.~(20).
This is because the gravitational Lagrangian has the usual form of a sum of terms homogeneous in ``velocities'' (i.e., time derivatives): $L_\mathrm{PG}=L_0+L_1+L_2$, consequently, from Euler's theorem, the associated {\em energy function} is of the familiar ``kinetic-minus-potential'' form:
$L_2-L_0$.  As Eq.~(\ref{quadcurv}) shows in detail, the ``new'' terms that we consider with parameters $w_2,w_4,\mu_4$ are---to order 2 and 0---\emph{identical} to those from $w_3,w_6,\mu_3$, respectively, so those parts \emph{contribute in exactly the same way},
while the order 1 terms do not effect the energy function.  Hence we have
\begin{eqnarray}
{\cal E}_\mathrm{PG}&:=&\frac{\partial L_\mathrm{PG}}{\partial \dot \psi}\dot\psi+\frac{\partial L_\mathrm{PG}}{\partial \dot\chi}\dot\chi+\frac{\partial L_\mathrm{PG}}{\partial \dot a}\dot a-L_\mathrm{PG}=L_2-L_0\nonumber\\
&=&a^3\biggl\{\Lambda-3a_0(H^2+ka^{-2})+\frac32 \tilde a_2(u^2-2Hu)
-6\tilde a_3x^2-6\tilde\sigma_2x(H-u)\nonumber\\
%&&
&&\qquad-\frac{w_{4+6}}{24}\left[R^2-12R\left\{(H-u)^2-x^2+ka^{-2}\right\}\right] \nonumber\\
&&\qquad
+\frac{w_{2+3}}{24}\left[X^2-24Xx(H-u)\right]\nonumber\\
&&\qquad-\frac{\mu_{3-2}}{24}\left[RX-6X\left\{(H-u)^2-x^2+ka^{-2}\right\}-12Rx(H-u)\right]\biggr\},\label{energyfunction}
\end{eqnarray}
where we introduced the convenient modified parameters
\begin{equation}
\tilde a_2:=a_2-2a_0,\quad \tilde a_3:=a_3-\frac12 a_0,\quad \tilde \sigma_2:=\sigma_2+b_0.
\end{equation}
The energy value~(\ref{energyfunction}) has the form $-a^3\rho$ where, from a comparison with Eq.~(167) in BHN~\cite{BHN}, it can be seen that $\rho$ is the value of the material {\em energy density}.

Since $L_{\rm PG}$ is time independent, the energy function~(\ref{energyfunction}) satisfies an {\em energy conservation} relation:
\begin{equation}
\dot{\cal E}_{\rm PG}
=-\frac{\delta L_{\rm PG}}{\delta \psi} {\dot \psi}-\frac{\delta L_{\rm PG}}{\delta \chi} {\dot \chi}-\frac{\delta L_{\rm PG}}{\delta a} {\dot a}=\frac{\delta L_{\rm int}}{\delta a}{\dot a}=3pa^2{\dot a},
\end{equation}
which, with the interpretation of $\rho$ mentioned above, is just the perfect fluid work-energy relation:
\begin{equation} -\frac{d(\rho a^3)}{dt}=p\frac{da^3}{dt}.\end{equation}

\subsection{Dynamical equations}
With the aid of the formulas for the torsion and curvature scalars
in terms of the gauge variables~(\ref{R}), (\ref{X}), (\ref{fchi}),
 the Euler-Lagrange equations for the gqPG cosmology
can be obtained. The $\psi$ equation is
\begin{eqnarray}
\frac{d}{dt}\frac{\partial L_\mathrm{PG}}{\partial \dot
\psi}&=&\frac{d}{dt}\left(\frac{a^{2}}{2}\left[6a_0-w_{4+6}R-\frac{\mu_{3-2}}2X\right]
-3w_{4-2}(\chi-\zeta)^2\right)\nonumber\\
&=&\frac{\partial L_\mathrm{PG}}{\partial \psi}\nonumber\\
&=&3(a_2u+2\sigma_2x)a^2\nonumber\\
&& +\left[6a_0-w_{4+6}R-\frac{\mu_{3-2}}2X
\right]a\psi \nonumber \\
&&+\left\{\left[6b_0-\frac{\mu_{3-2}}2 R+w_{3+2}X\right]a+6w_{4-2}\dot\chi\right\}(\chi-\zeta).\label{Epsi}
\end{eqnarray}
Reexpressed in terms of observables $(R,\ X, u,\ x, H)$ the $\psi$ equation becomes
\begin{eqnarray}
-\frac{w_{4+6}}2\dot R-\frac{\mu_{3-2}}4 \dot
X&=&-\left[-3 \tilde a_2-w_{4+6}R-\frac{\mu_{3-2}}2
X\right]u\nonumber\\
&& +\left[6\tilde\sigma_2-\frac{\mu_{3-2}}2 R+w_{2+3}X\right]x\nonumber\\
&&+w_{4-2}[2X-24(H-u)x]x. \label{dotRdotX} %\label{Epsi}
\end{eqnarray}
The last line includes qualitatively new types of terms when compared to the corresponding result in the earlier investigations~\cite{HN,AdP} based on the BHN model~\cite{BHN}.

The $\chi$ equation is
\begin{eqnarray}
\frac{d}{dt}\frac{\partial L_\mathrm{PG}}{\partial \dot
\chi}&=&\frac{d}{dt}\left\{\frac{a^{2}}{2}\left[6b_0-\frac{\mu_{3-2}}2R+w_{3+2}X \right]+6w_{4-2}\psi(\chi-\zeta)\right\}\nonumber\\
&=&\frac{\partial L_\mathrm{PG}}{\partial \chi}\nonumber\\
&=&6(2a_3x-\sigma_2u)a^2\nonumber\\
&& -\left\{\left[6a_0-w_{4+6}R
-\frac{\mu_{3-2}}2
X\right]a+6w_{4-2}\dot\psi\right\}(\chi-\zeta) \nonumber \\
& &+\left\{\left[6b_0-\frac{\mu_{3-2}}2 R+w_{2+3}X\right]a+6w_{4-2}\dot\chi\right\}\psi.\label{Echi}
\end{eqnarray}
Reexpressed in terms of observables this becomes %$\Longrightarrow$
\begin{eqnarray}
-\frac{\mu_{3-2}}4 \dot
R+\frac{w_{2+3}}2\dot X&=&-\left[6\tilde\sigma_2-\frac{\mu_{3-2}}2 R+w_{2+3}X\right]u\nonumber\\
&&+\left[12\tilde a_3+w_{4+6}R+\frac{\mu_{3-2}}2
X\right]x\nonumber\\
&&-w_{4-2}(2R-12[(H-u)^2-x^2+k a^{-2}])x. \label{dotXdotR} %\label{Echi}
\end{eqnarray}
Again the last line shows novel interaction terms compared to those seen in the earlier BHN model investigations.

For the $a$ equation, however, the result has exactly the same form as that found in the earlier BHN model investigations~\cite{HN,AdP} since, by~(\ref{quadcurv}), the $w_{4-2}$ terms are independent of the scale factor $a$, and so make no contribution:
\begin{eqnarray}
\frac{d}{dt}\frac{\partial L_\mathrm{PG}}{\partial \dot
a}&=&\frac{d}{dt}\left(-a^{2}3[a_2u+2\sigma_2x]\right)=\frac{\partial L_\mathrm{PG}}{\partial a}
      +\frac{\partial L_\mathrm{int}}{\partial a}\nonumber\\
&=&3a^{-1}L-\left(\frac{a_0}2-\frac{w_{4+6}}{12}R-\frac{\mu_{3-2}}{24}X\right)[a^2R+6(\psi^2-[\chi-\zeta]^2+\zeta^2)]\nonumber\\
&&-\left(\frac{b_0}2+\frac{w_{2+3}}{12}X-\frac{\mu_{3-2}}{24}R\right)[a^2X+12\psi(\chi-\zeta)]\nonumber\\
&&+3a^2(a_2u+2\sigma_2x)u-6a^2[2a_3x-\sigma_2u]x+3pa^2.
\end{eqnarray}
Reexpressed in terms of observables this takes the form
\begin{eqnarray}
-3(a_2\dot u+2\sigma_2 \dot x)
 &=&
a_0R+b_0X-3\Lambda+3p+6H[a_2u+2\sigma_2x]\nonumber\\
&&+\frac3{2}(-a_2u^2+4a_3x^2-4\sigma_2ux)\nonumber\\
&&+\frac{a_0}2\left[-6(H-u)^2+6x^2-6ka^{-2}\right]
-6b_0x(H-u)\nonumber
\\
&&-\frac{w_{4+6}}{24}\{R^2-12R[(H-u)^2-x^2+ka^{-2}]\}\nonumber\\
&&-\frac{\mu_{3-2}}{24}\{RX-12R(H-u)x-6X[(H-u)^2-x^2+ka^{-2}]\}\nonumber\\
&&+\frac{w_{2+3}}{24}\{X^2-24X(H-u)x\}.\label{Ea}
\end{eqnarray}

Two further useful relations can be obtained by taking the time
derivatives of the torsion~(\ref{fchi}) and using the curvature definitions~(\ref{R},\ref{X}):
\begin{eqnarray}
\qquad\dot x&=&\frac{X}6-Hx-2x(H-u),\label{xdot}\\
\dot H -\dot u&=&\frac{R}6-H(H-u)-[(H-u)^2-x^2+ka^{-2}]\label{Hudot}.
\end{eqnarray}

Rearranging~(\ref{Ea}) using (\ref{xdot}) yields
\begin{eqnarray}
-3a_2\dot u &=&
a_0\{R-3[(H-u)^2-x^2+ka^{-2}]\}-3\Lambda+3p\nonumber\\
&&+\tilde\sigma_2[X-6(H-u)x] %\nonumber\\
%&&
+\frac3{2}(-a_2u^2+4a_3x^2)+6a_2Hu\nonumber\\
&&-\frac1{24}w_{4+6}\{R^2-12R((H-u)^2-x^2
+ka^{-2})\}\nonumber\\
&&-\frac1{24}\mu_{3-2}\{RX-12R(H-u)x-6X[(H-u)^2-x^2+ka^{-2}]\}\nonumber\\
&&+\frac1{24}w_{2+3}[X^2-24X(H-u)x]. \label{udot3p}
\end{eqnarray}
Using the expression for the material energy density gives a more compact alternative:
\begin{eqnarray}
3a_2\dot u &=&
-a_0R-\tilde\sigma_2X+\rho-3p+4\Lambda\nonumber\\
&&+3a_2(u^2-3Hu)-12a_3x^2. \label{udotrho3p}
\end{eqnarray}

\subsection{Alternative derivations}

The system of dynamical equations can also be obtained from simpler Lagrangians that can be found using the topological identities~(\ref{euler},\ref{pointryagin}). These alternatives serve as good checks of the calculation.

One ``minimal'' Lagrangian has the previously studied BHN form plus the $w_4$ term.
This additional quadratic term can be rewritten as follows:
\begin{eqnarray}
a^3\tilde R^2&\equiv& a^3(\tilde R^2-R^2)+a^3R^2\nonumber\\
& \equiv& -4\cdot6^2\dot\psi[\psi^2-(\chi-\zeta)^2+\zeta^2]+a^3 R^2\nonumber\\
&\equiv& a^3R^2 +4\cdot 6^2 \dot\psi(\chi-\zeta)^2 + \hbox{ total time derivative}.
\end{eqnarray}
This leads to ``extra'' terms in the $\psi$ and $\chi$ equations which are equivalent to those found above.

Furthermore, one could instead use the Euler identity to eliminate the $w_4$ term.  This leads to another ``minimal'' version with the effective Lagrangian being BHN plus the $w_2$ term.  The additional quadratic curvature term can be written as
\begin{eqnarray}
a^3\tilde X^2&\equiv& a^3X^2+a^3(\tilde X^2-X^2)\nonumber\\
&\equiv& a^3X^2 -8\cdot6^2 \dot\chi\psi(\chi-\zeta).
\end{eqnarray}
This leads to the same ``extra'' terms in the $\psi$ and $\chi$ equations as found above.

\section{First order equations}

The above equations~(\ref{Epsi}), (\ref{Echi}), and (\ref{Ea}) are 3 {\em second order} equations for the \emph{gauge potentials} $\psi,\chi,a$.
However they can be used to construct a set of 6 {\em first order} equations for the \emph{observable} quantities:
namely~(\ref{dotRdotX}), (\ref{dotXdotR}), (\ref{xdot}), (\ref{Hudot}), and (\ref{udot3p}) or (\ref{udotrho3p}) along with the Hubble relation.

For convenience we repeat them together here:
\begin{eqnarray}
-\frac{w_{4+6}}2\dot R-\frac{\mu_{3-2}}4 \dot
X&=&-\left[-3 \tilde a_2-w_{4+6}R-\frac{\mu_{3-2}}2
X\right]u\nonumber\\
&& +\left[6\tilde\sigma_2-\frac{\mu_{3-2}}2 R+w_{2+3}X\right]x\nonumber\\
&&+w_{4-2}[2X-24(H-u)x]x, \label{dotRdotX'} %\label{Epsi}
\\
-\frac{\mu_{3-2}}4 \dot
R+\frac{w_{2+3}}2\dot X&=&-\left[6\tilde\sigma_2-\frac{\mu_{3-2}}2 R+w_{2+3}X\right]u\nonumber\\
&&+\left[12\tilde a_3+w_{4+6}R+\frac{\mu_{3-2}}2
X\right]x\nonumber\\
&&-w_{4-2}(2R-12[(H-u)^2-x^2+k a^{-2}])x, \label{dotXdotR'} %\label{Echi}
\end{eqnarray}
%\\
\begin{eqnarray}
\dot a&=&aH, \label{hubble'}\\
%\begin{eqnarray}
\qquad\dot x&=&\frac{X}6-3Hx+2xu,\label{xdot'}\\
\dot H -\dot u&=&\frac{R}6-2H^2+3Hu-u^2+x^2-ka^{-2},\label{Hudot'}
\\
a_2\dot u &=&\frac13(
-a_0R-\tilde\sigma_2X+\rho-3p+4\Lambda)\nonumber\\
&&+a_2(u^2-3Hu)-4a_3x^2. \label{udotrho3p'}
\end{eqnarray}

The associated material energy density is
\begin{eqnarray}
\rho
&=&-\Lambda+3a_0[(H-u)^2-x^2+ka^{-2}]\nonumber\\&&-\frac32 a_2(u^2-2Hu)
+6 a_3x^2+6\tilde\sigma_2x(H-u)\nonumber\\
&&\qquad+\frac{w_{4+6}}{24}\left[R^2-12R\left\{(H-u)^2-x^2+ka^{-2}\right\}\right] \nonumber\\
&&\qquad+\frac{\mu_{3-2}}{24}\left[RX-6X\left\{(H-u)^2-x^2+ka^{-2}\right\}-12Rx(H-u)\right]\nonumber\\
&&\qquad
-\frac{w_{2+3}}{24}\left[X^2-24Xx(H-u)\right]\label{density}.
\end{eqnarray}

The two dynamical equations~(\ref{dotRdotX'},\ref{dotXdotR'})
can be resolved for $\dot R$ and $\dot X$ by inverting the symmetric matrix
\begin{equation}
\frac14\left(
  \begin{array}{cc}
    -2w_{4+6} & -\mu_{3-2} \\
    -\mu_{3-2} & 2w_{2+3} \\
  \end{array}
\right). \label{kineticT}
\end{equation}
The qualitatively ``new'' terms appear only in the second lines of these 2 relations.
 These ``new'' terms exclude a special class of solutions with certain constant values for the scalar curvatures $R$ and $X$.  When $w_{4-2}$ vanishes we are back at the BHN case investigated earlier~\cite{BHN,HN,AdP}.  That model, with one further parameter restriction, admits a certain class of constant curvature solutions~\cite{HN15}.

It should be noted that we have written our equations in a form in which they are valid for all ranges of the parameters, including both the ``physical'' and ``unphysical'' ranges, as well as degenerate cases.  Generically there are 3 degrees of freedom, but in certain degenerate cases there are less, such as when $a_2$ vanishes or when the matrix~(\ref{kineticT}) has less than maximal rank.  Then some of the above equations change their character.  They may become constraint equations rather than dynamical equations.  A particularly interesting subclass is the generalized Einstein-Cartan system, wherein all the quadratic curvature parameters vanish; this includes GR as a special case.  An analysis of that subclass is given in Ref.~\cite{HN15}.  From here on we confine our attention the \emph{generic} case, with no special degeneracies.

%%%%%%%%%%%%%%%%%%%%%%%%%%%%%%%%%%%%%%%%%%%%%%%
\section{Linearized equations and late time normal modes}\label{linana}
%%%%%%%%%%%%%%%%%%%%%%%%%%%%%%%%%%%%%%%%%%%%%%%
 The fluid pressure $p$, the cosmological constant $\Lambda$, and the spatial curvature parameter $k$ play essentially the same roles here in the gqPG as they play in the GR FLRW cosmologies.  Hence, in order to better see the novel features of gqPG cosmology, we take them to vanish in our further considerations.

Following the procedure used in~\cite{JCAP09,AdP}, {\em for vanishing $\Lambda$ and $k$}, by dropping higher than linear order terms in $H$, $u$, $x$, $R$, $X$ in the dynamical equations, our model can be linearized. (It should be noted that under these assumptions $\rho$ and hence $p$ are non-linear.) This leads to the first order {\em linearized\/} versions of the dynamical equations.

Using this procedure for the BHN model one zero frequency and two dynamical normal modes were identified.  Furthermore, it was argued that the dynamical variables would have a fall off of $a^{-3/2}$. Hence the dynamical variables rescaled by $a^{3/2}$ would at late time just approach the linearized modes.  This behavior was confirmed by numerical calculations~\cite{AdP}.

A comparison of the gqPG equations considered here with those of the BHN model considered in~\cite{AdP} shows that the new terms that appear in the general system, those with the coefficient $w_{4-2}$, are all \emph{non-linear}.  Hence the \emph{linearization} of the gqPG is essentially identical with that of the BHN model, and thus the gqPG cosmology has essentially the same linearized normal modes and frequencies as were found earlier for the BHN model.
 Consequently the ``late time'' normal mode analysis for the $\Lambda=0=k$ case (as discussed in~\cite{AdP}) applies to general quadratic PG cosmology virtually unchanged.
That the gqPG really does show essentially these same qualities has been demonstrated in our numerical analysis described in the next section.

We note that the zero frequency mode for the gqPG is
\begin{equation}
z:= a_0(H-u)+\frac{{a}_2}{2}u+\tilde{\sigma}_2x,\label{zero}
\end{equation}
(from Eqs.~(\ref{dotRdotX'})--(\ref{udotrho3p'}) it can be verified that this is indeed constant to linear order)
while the details of the normal modes and frequencies are, as mentioned, essentially identical with those presented in~\cite{AdP}.

\section{numerical}
%%%%%%%%%%%%%%%%%
% Numerical parts by Hsin Chen and Feihung 8/23
%%%%%%%%%%%%%%%%%

\begin{figure}[thbp]\label{fig1}
\begin{tabular}{rl}
\includegraphics[width=0.5\textwidth]{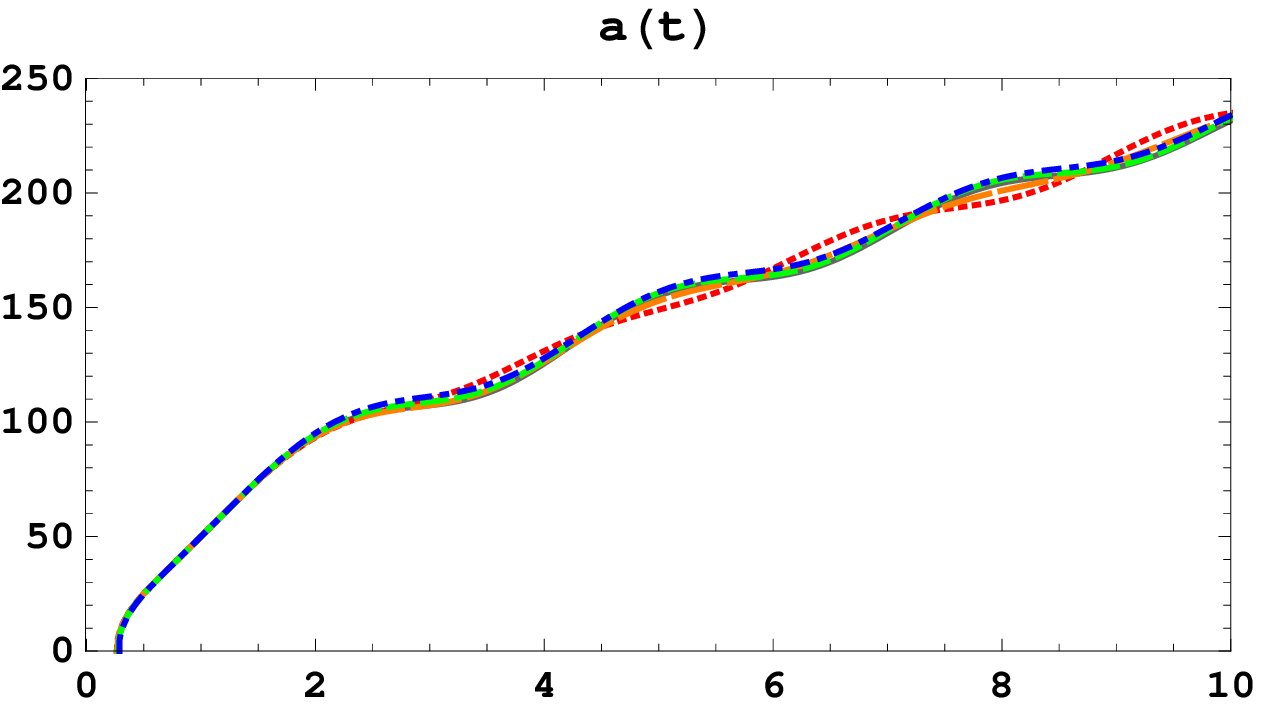}%&
\includegraphics[width=0.5\textwidth]{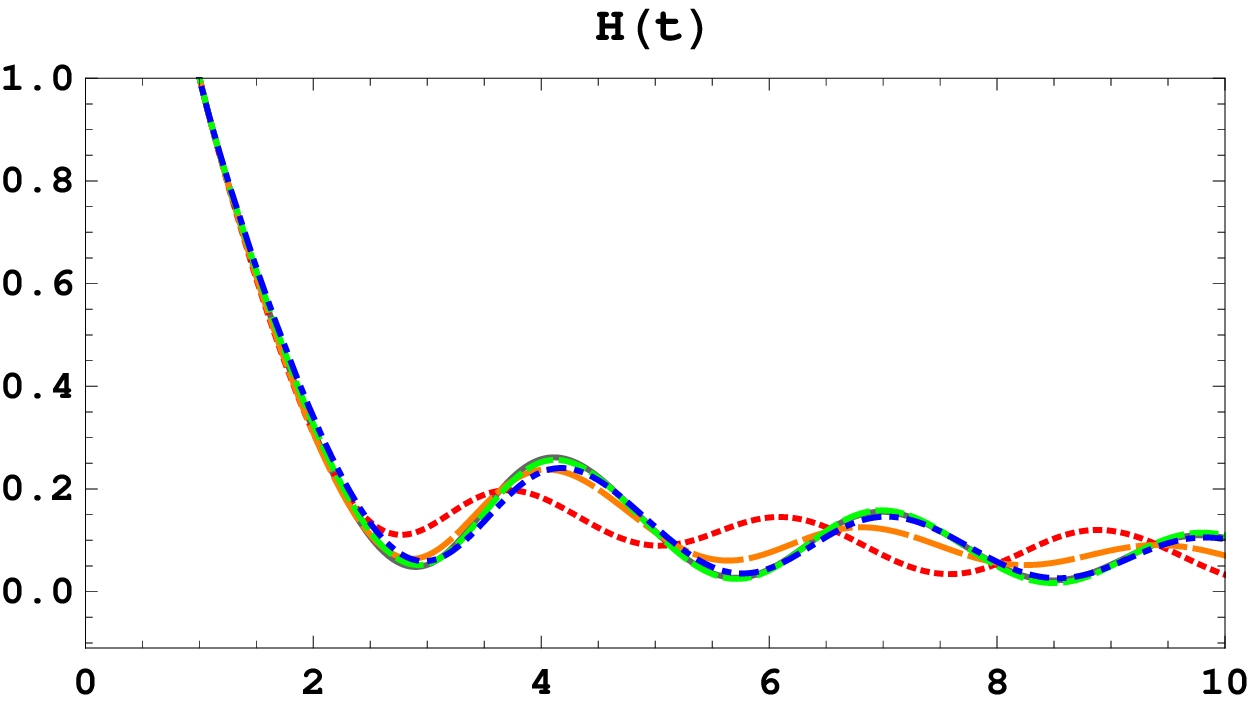}\\
\includegraphics[width=0.5\textwidth]{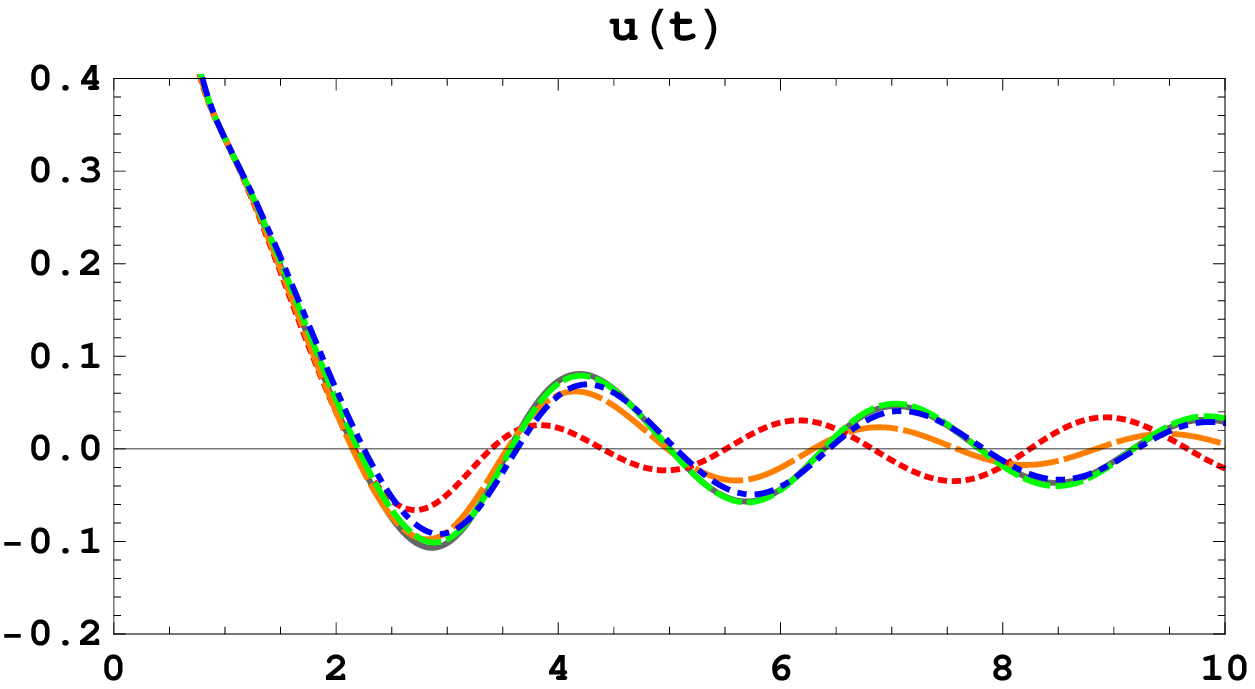}%&
\includegraphics[width=0.5\textwidth]{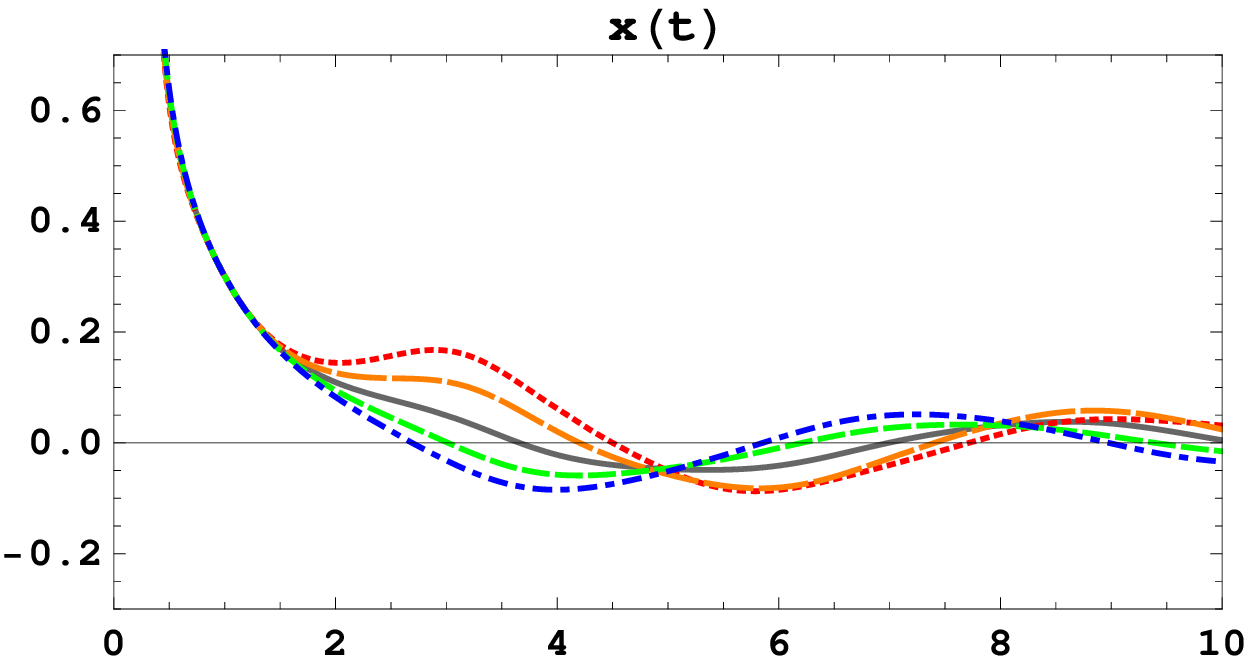}\\
\includegraphics[width=0.5\textwidth]{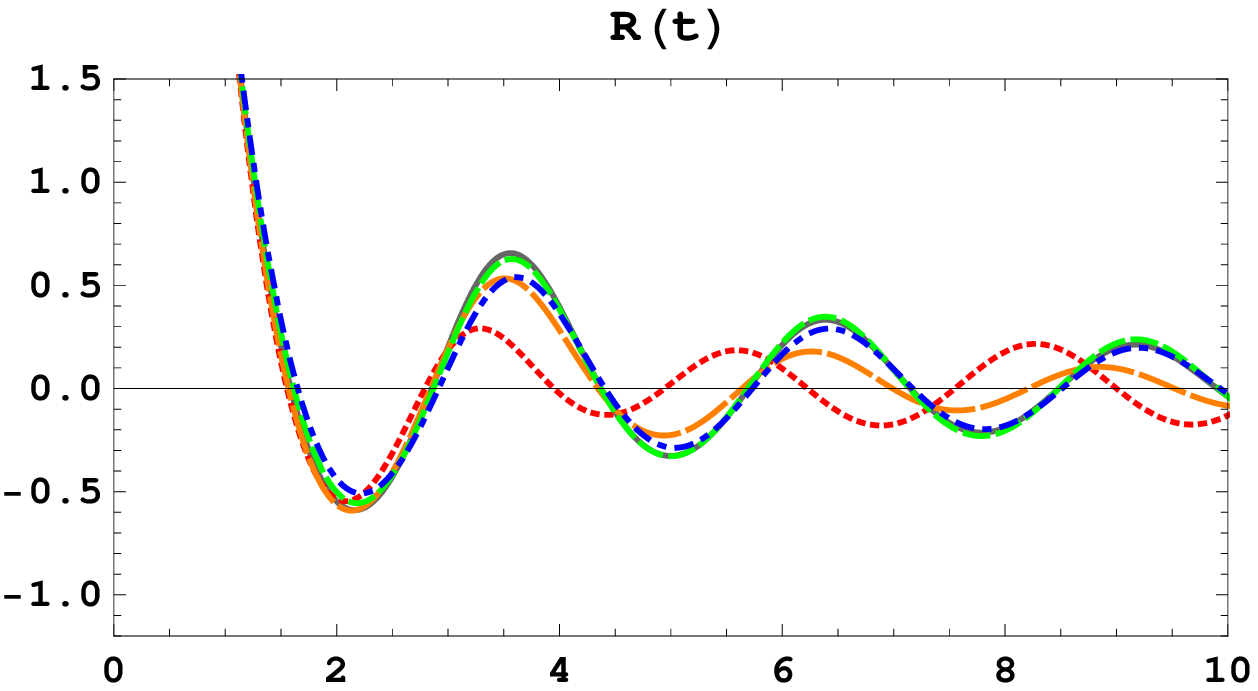}%&
\includegraphics[width=0.5\textwidth]{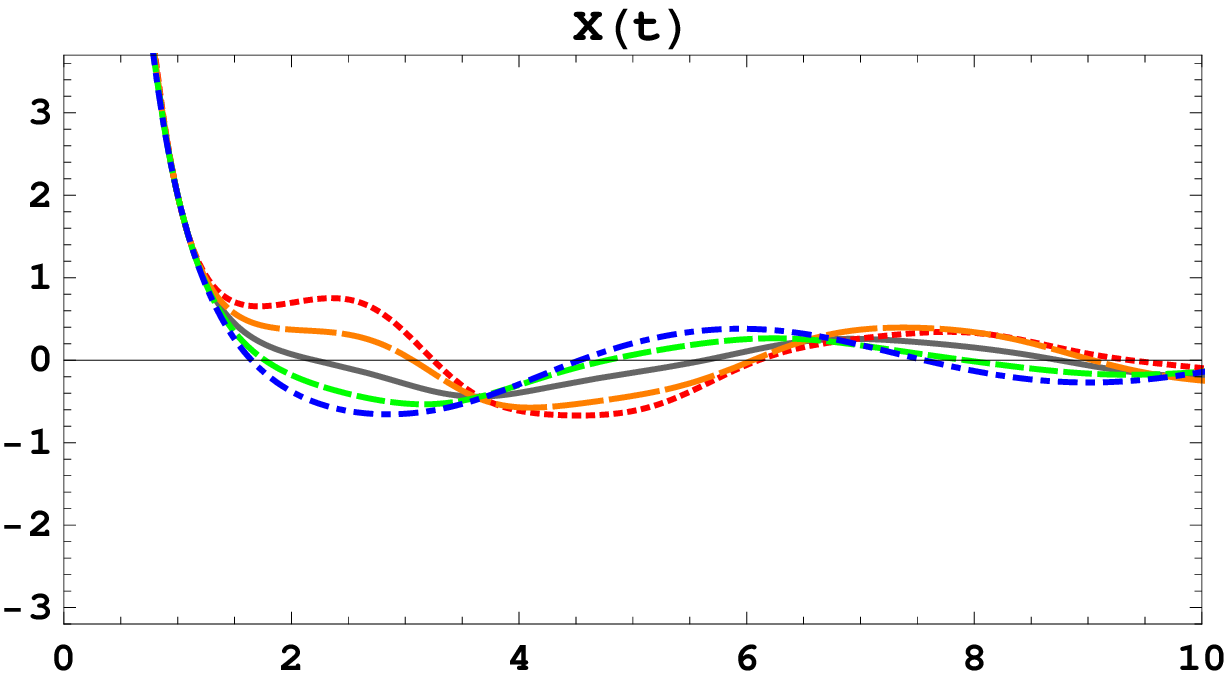}\\
\includegraphics[width=0.5\textwidth]{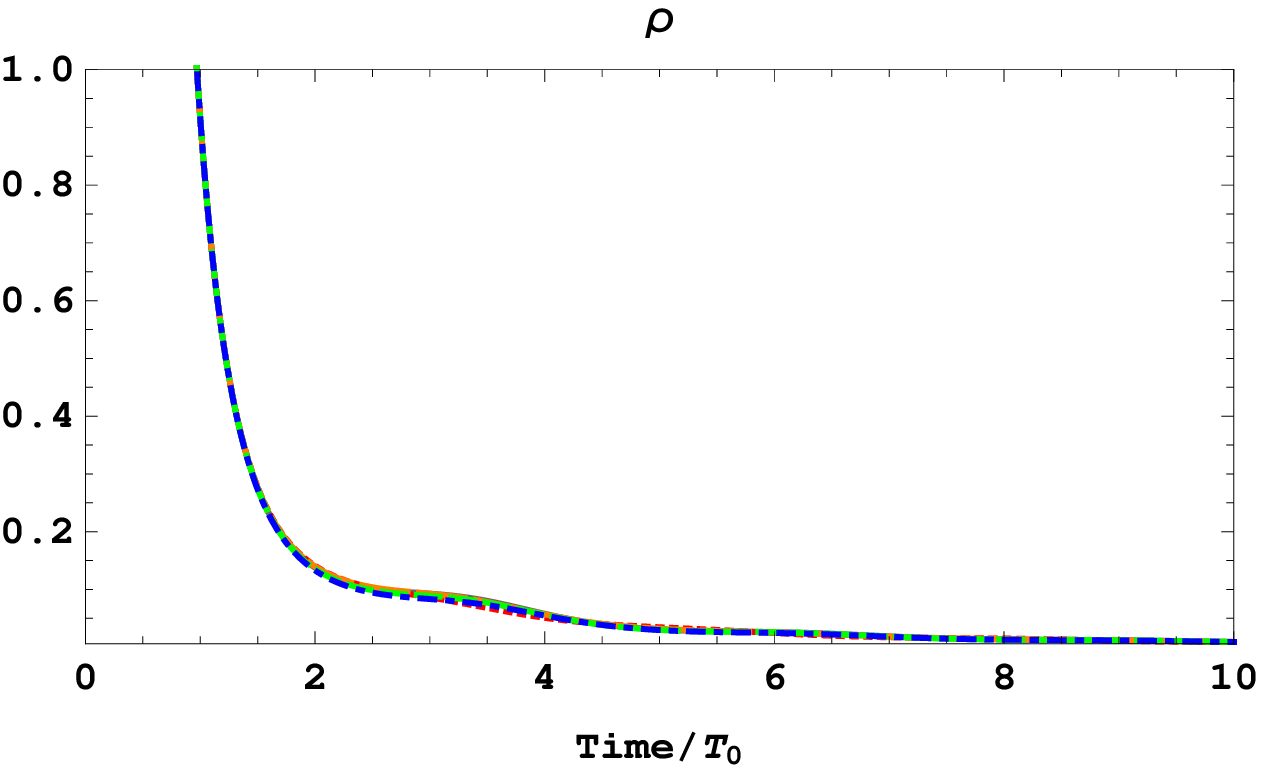}%&
\includegraphics[width=0.5\textwidth]{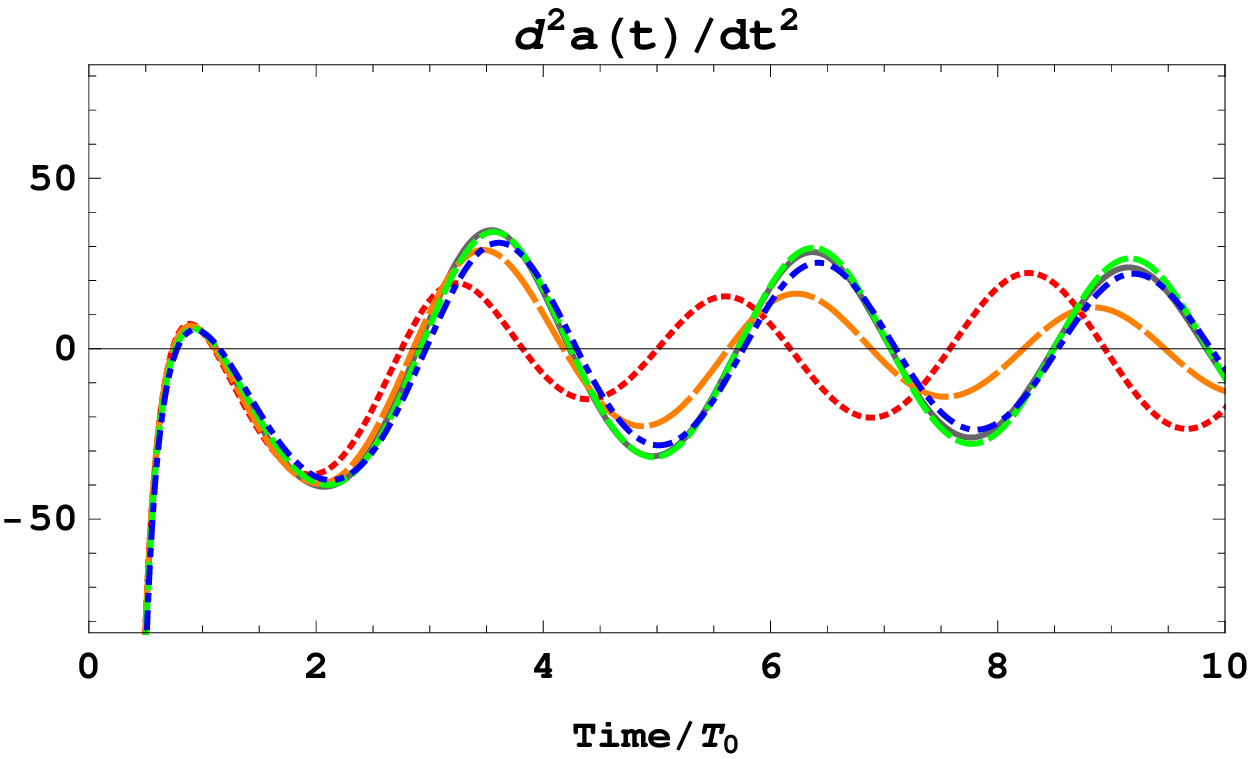}%\\
\end{tabular}
\caption{The dynamics of the observables $a,\ H,\ u,\ x,\ R,\ X,\ \rho,\ \ddot a$, showing the
effects of the parameter $w_{4-2}$. The red tiny dashed line is for $w_{4-2}=1$, the orange long dashed line is for $w_{4-2}=0.5$, the gray solid line is for $w_{4-2}=0$, the green medium dashed line line is for $w_{4-2}=-0.5$, and the blue dot-dashed is for $w_{4-2}=-1$.}
\end{figure}

For the BHN model the effects of the PG pseudoscalar parameters $\mu_{3},\ \tilde\sigma_2$ that algebraically cross couple the spin $0^+$ and $0^-$ modes had been investigated earlier~\cite{HN}. The linearized equations and the associated normal modes and their frequencies were found~\cite{AdP}.  It was noted that (with vanishing $p,\ \Lambda,\ k$) from the conserved energy condition $a^3\rho=$const in an expanding universe one could infer that the observables $H,\ u,\ x,\ R,\ X$ rescaled by $a^{3/2}$ would at late time satisfy the linearized equations.  This was supported by numerical evolution that indeed showed the expected damping and the various observables as superpositions of two normal modes and one zero frequency mode.

Likewise for the gqPG FLRW cosmology, the parameters $\tilde\sigma_2,\ \mu_{3-2}$ will play the same kind of role, and again (with vanishing $p,\ \Lambda,\ k$) from the conserved energy condition $a^3\rho=$const one can infer that the observables rescaled by $a^{3/2}$ should at late time behave like the linearized equation solutions.

To verify this expectation and to see the effects of the new parameter $w_{4-2}$
we follow the numerical analysis technique used previously~\cite{SNY08,JCAP09,HN,AdP}.
 Based on certain sets of well-tested appropriate parameters, typical effects of the new parameter $w_{4-2}$ can be seen from the parameter choices
\begin{eqnarray}
&&a_0=1; \quad a_{2}=-0.85;\quad a_{3}=0.35;\quad w_{4+6}=-1.3; \quad w_{2+3}=0.6; \nonumber\\
&&\tilde\sigma_2=0;\quad \mu_{3-2}=0; \quad w_{4-2}=1,\ 0.5,\ 0,\ -0.5,\ -1.
\end{eqnarray}
We have chosen the cross parity pseudoscalar coupling parameters (whose effects were explored in~\cite{HN}) to vanish here in order to more clearly see the effects of the ``new'' cross parity coupling $w_{4-2}$.
The initial values we used are
\begin{eqnarray}
&&a(t_0)=50,\quad H(t_0)=1,\quad u(t_0)=0.335,\quad x(t_0)=0.3, \nonumber\\
&&R(t_0)=2.18, \quad X(t_0)=2.0,
\end{eqnarray}
where $t_0=1$, the present time of universe.

Fig.~1 displays typical evolutions for the observables $a,\ H,\ u,\ x,\ R,\ X,\ \rho,\ \ddot a$, showing how the behavior depends on the ``new'' parameter $w_{4-2}$.
For the same parameter values and initial conditions the rescaled variables $a^{3/2}H$, $a^{1/2}\textrm{d}^2a/\textrm{d}t^2$, $a^{3/2}R$, $a^{3/2}X$, $a^{3/2}u$, $a^{3/2}x$, $a^{3/2}z$, $a^3\rho$ are plotted in Fig.~2.

From Fig.~1 one can see the typical damped oscillation behavior for $H,\ u,\ x,\ R,\ X$ that is obtained with generic parameter values in the physical range.  Note that the Hubble parameter $H$ is decreasing on average, but in repeating cycles is sometimes increasing.  This is reflected in the acceleration $\ddot a$.  On average the expansion is slowing down but has repeating cycles of acceleration.

From Fig~2 one can readily see that the amplitude for the oscillating observables rescaled by $a^{3/2}$ indeed is tending to a constant at late time.  The nearly constant value of $a^3\rho$ is a good indication of the numerical stability accuracy. The plot of $a^{3/2}z$ is a good indicator that, for this set of parameters and initial values, before $t=10$ the linearized approximation is not so good but it is not bad after $t=20$.
We note a rather dramatic change in these waveforms when $w_{4-2}$ is changed from 1 to 0.5 to 0, but a relatively smoother change in the range 0 to $-0.5$ to $-1$. As expected, for rescaled $H,\ \ddot a,\ R,\ X,\ u,\ x$ the waveforms are approximately harmonic, with nearly uniform amplitude and wavelength. This is especially the case  for $w_{4-2}=0$ (which is equivalent to the BHN model) except at the very earliest times. For nonvanishing values of the parameter $w_{4-2}$ these observables are not very harmonic at early times, such as $t<10$.  But at later times, like $t>15$, they are quite harmonic. Thus, as expected, the parameter $w_{4-2}$ can have a large effect at early time, but it does not have much effect at later time.  Whatever wave pattern it produced is at late time preserved.

These features are just what we expected from an examination of the $w_{4-2}$ terms in the dynamical equations.  From~(\ref{dotRdotX}) and (\ref{dotXdotR}) it is clear that the $w_{4-2}$ terms are nonlinear, and in an expanding universe they will at late time have no effect.  The $w_{4-2}$ terms couple the two dynamical modes of even and odd parity in a quite different fashion than that associated with the parameters $\mu_{3-2},\ \tilde\sigma_2$.  The latter are \emph{pseudoscalar} parameters that produce a linear coupling of the modes.  But $w_{4-2}$ is a \emph{scalar} parameter controlling a non-linear coupling of a kind of magnetic type, as revealed by the $w_{4-2}$ terms in Eqs.~(\ref{Epsi}) and (\ref{Echi}) which have the form of a product of a gauge variable with the time derivative of a gauge variable.

\begin{figure}[thbp]\label{fig2}
\begin{tabular}{rl}
\includegraphics[width=0.5\textwidth]{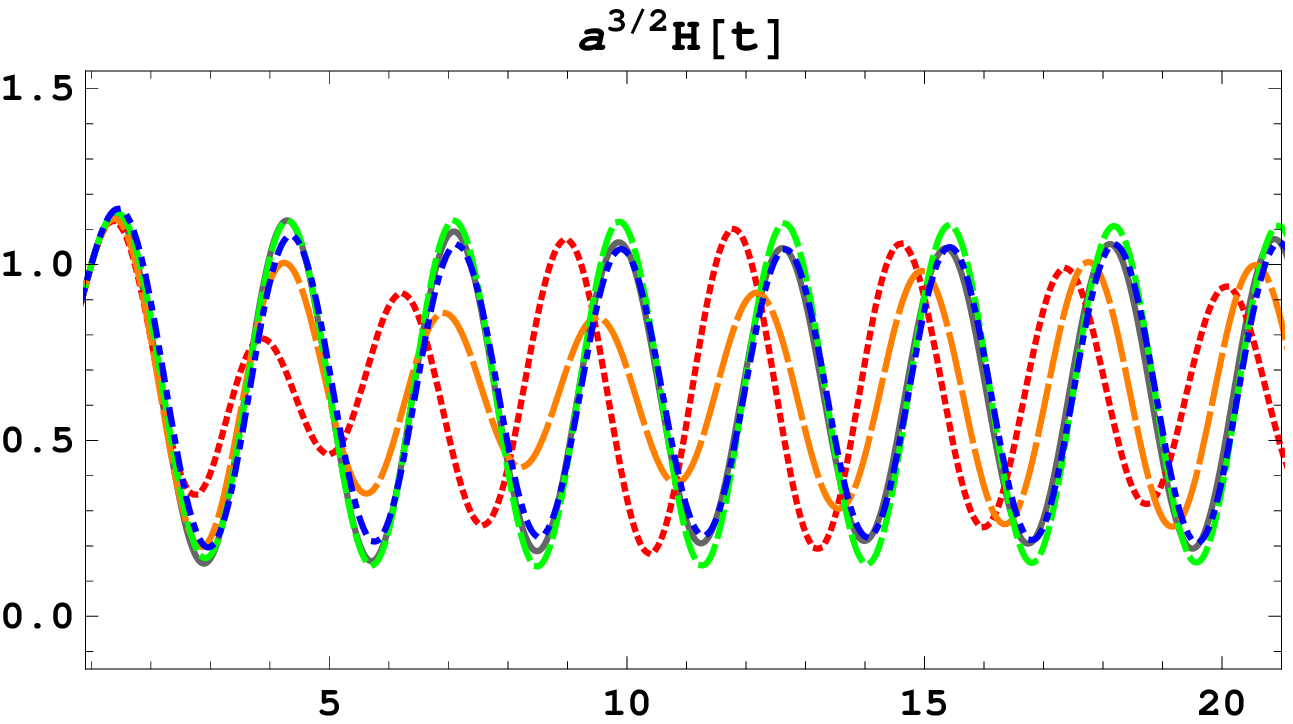}%&
\includegraphics[width=0.5\textwidth]{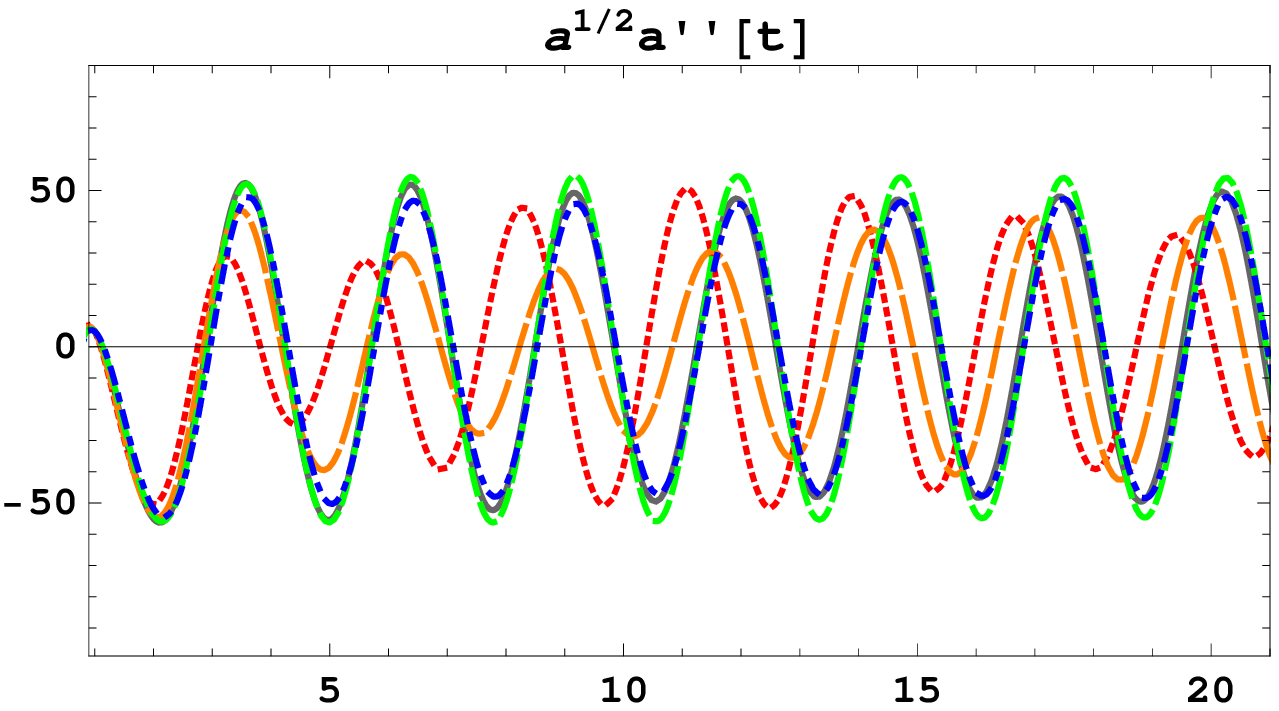}\\
\includegraphics[width=0.5\textwidth]{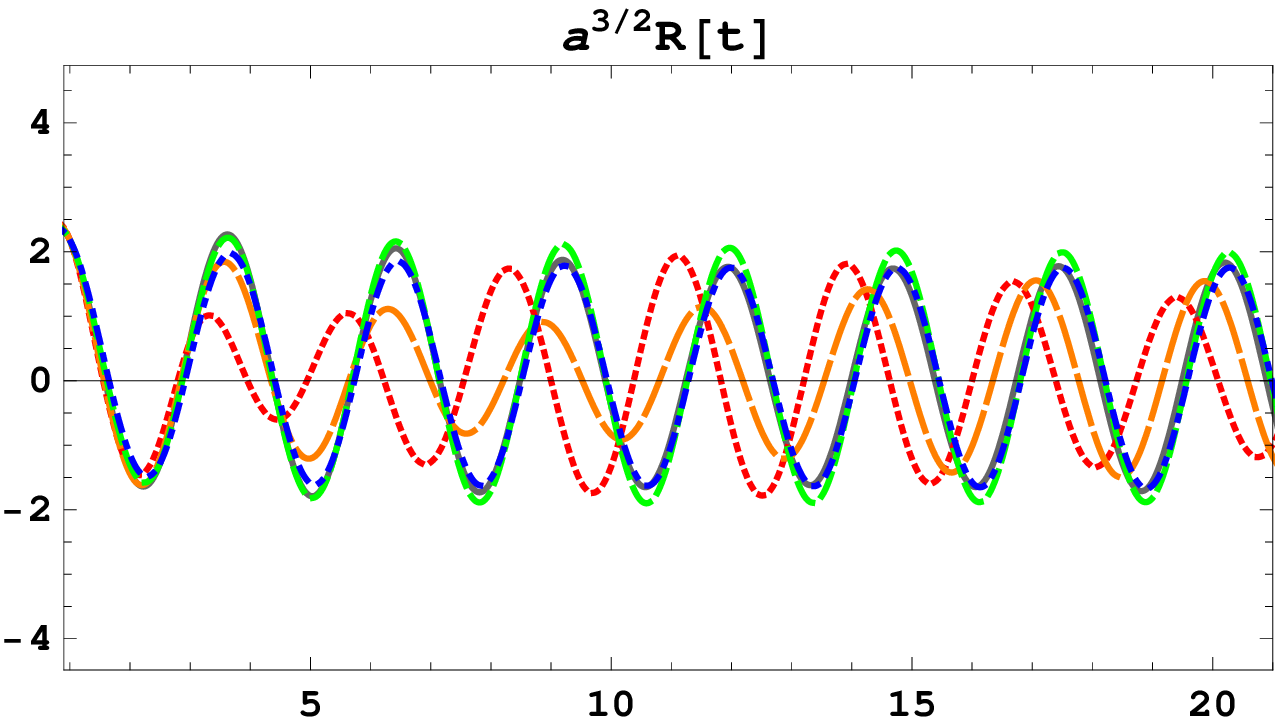}%&
\includegraphics[width=0.5\textwidth]{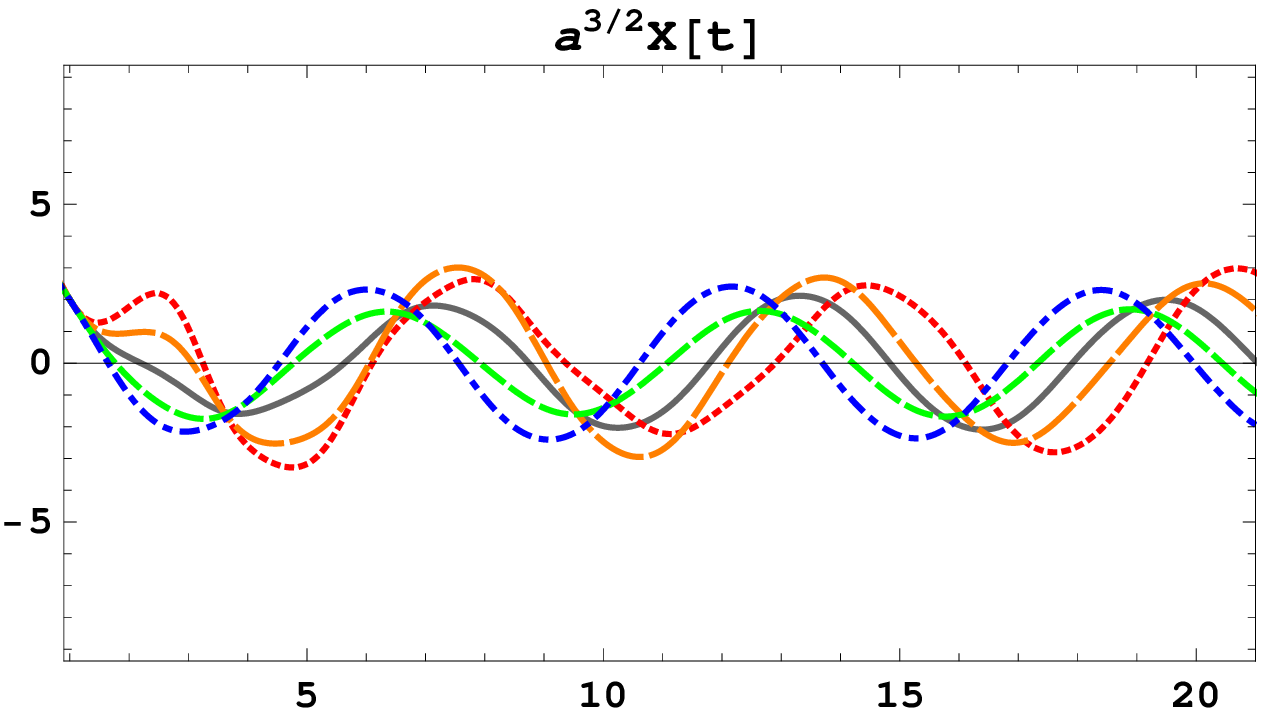}\\
\includegraphics[width=0.5\textwidth]{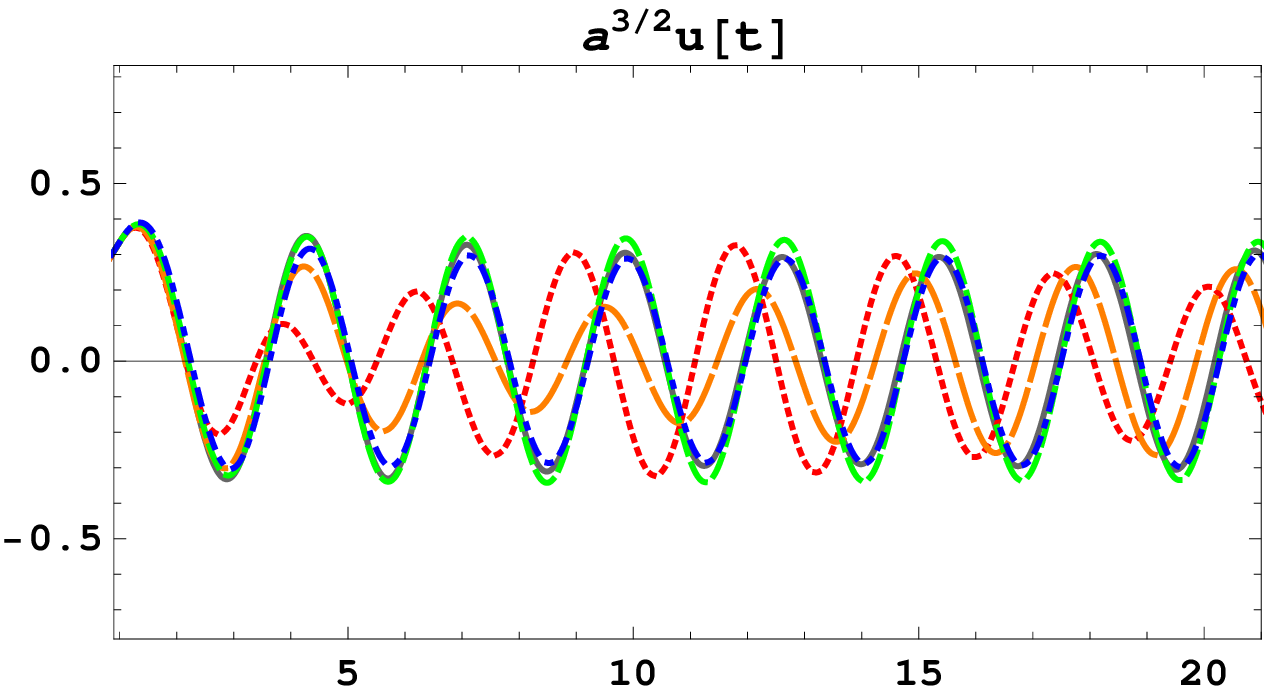}%&
\includegraphics[width=0.5\textwidth]{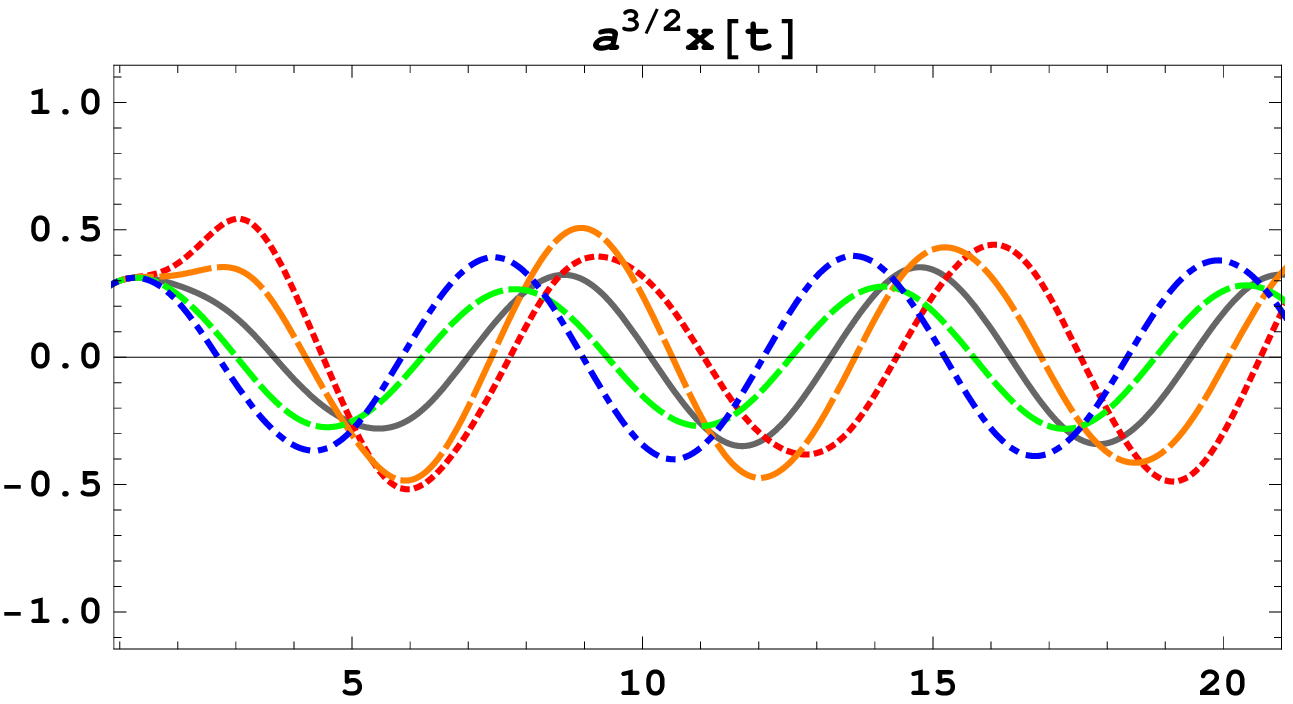}\\
\includegraphics[width=0.5\textwidth]{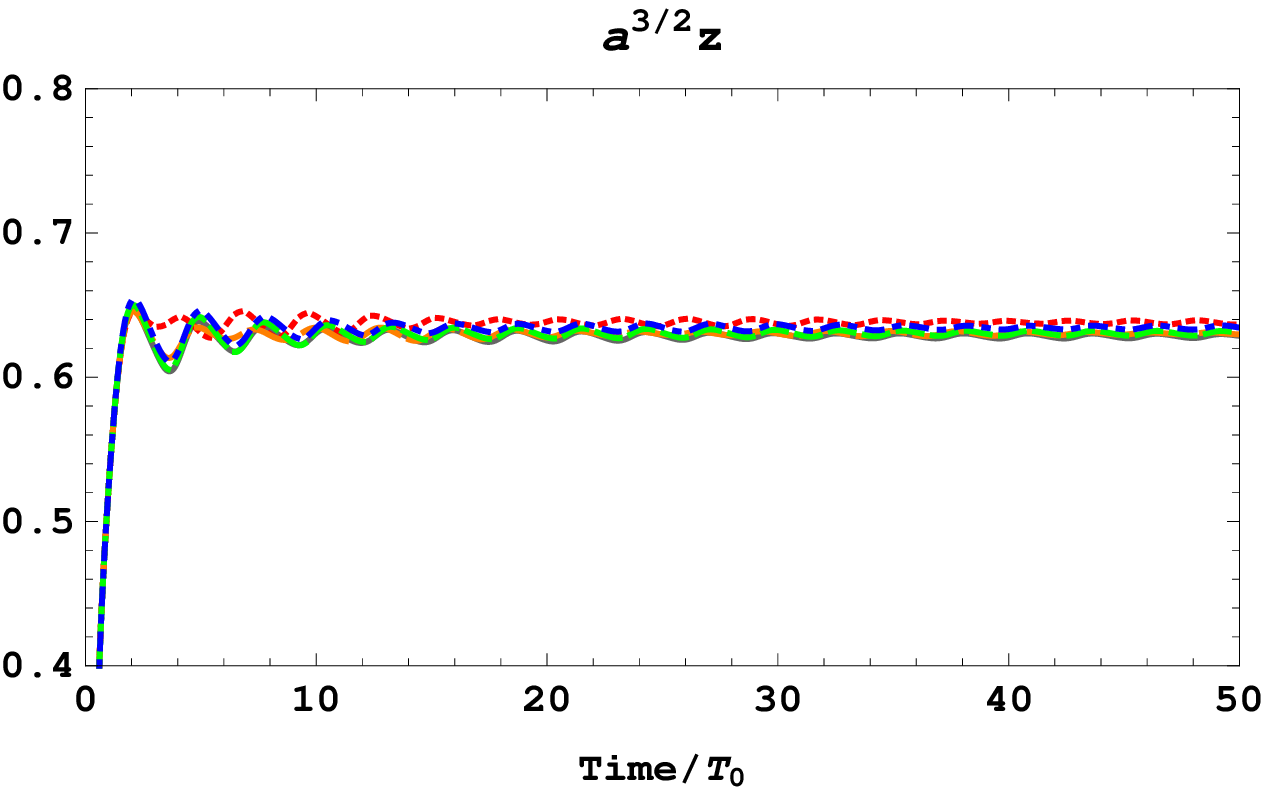}%&
\includegraphics[width=0.5\textwidth]{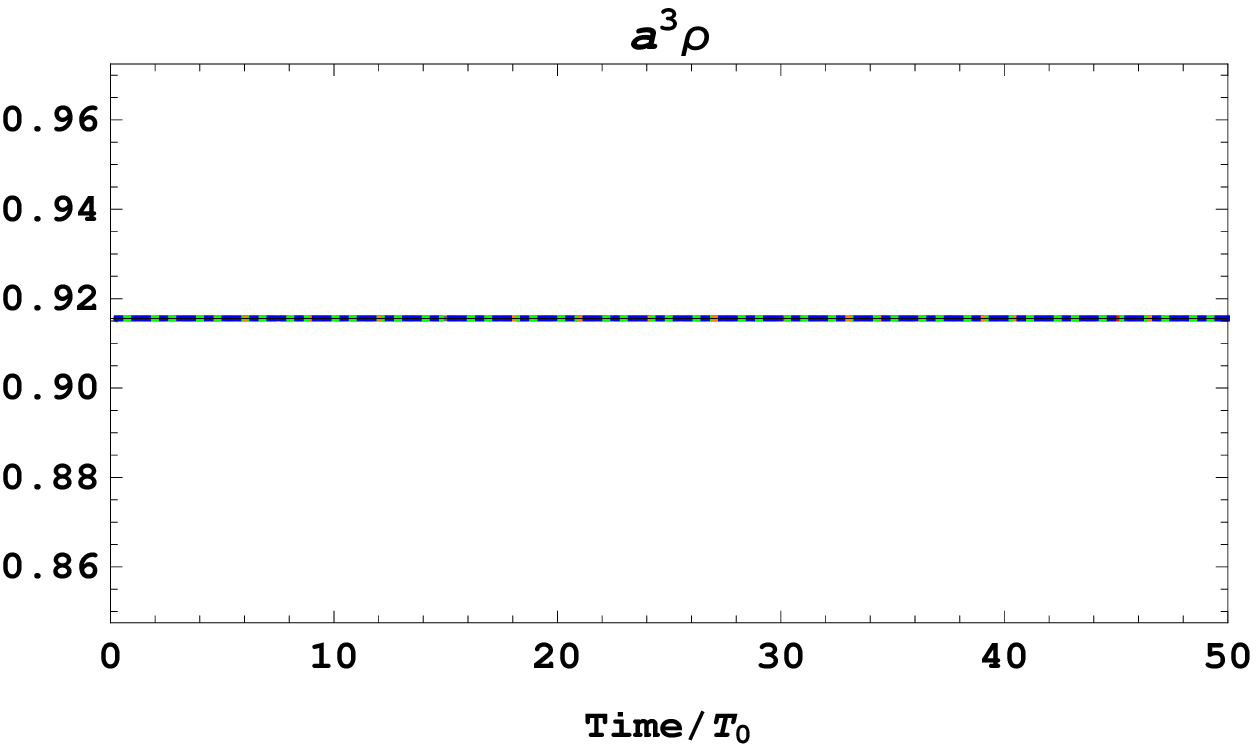}
\end{tabular}
\caption{The effects of the new parameter $w_{4-2}$ on the rescaled observables $a^{3/2}H,\ a^{3/2}\ddot a,\ a^{3/2} R,\ a^{3/2}X,\ a^{3/2} u,\ a^{3/2}x,\ a^{3/2} z,\ a^3\rho$.  The red tiny dashed line is for $w_{4-2}=1$, the orange long dashed line is for $w_{4-2}=0.5$, the gray solid line is for $w_{4-2}=0$, the green medium dashed line line stand is $w_{4-2}=-0.5$, and the blue dot-dashed is for $w_{4-2}=-1$.}
\end{figure}

\section{Discussion}

Here we have considered FLRW cosmological models for the general quadratic PG, including all the possible even and odd parity terms with their respective scalar and pseudoscalar parameters.
For the general PG all six connection modes are in principle dynamic---however the cosmological symmetries essentially only allow \emph{scalar} dynamic connection modes.
 Generically the general quadratic PG cosmology has, in addition to the usual metric scale factor, just two dynamical connection modes with spin-$0$: $0^+$ and $0^{-}$, effectively a scalar and a pseudoscalar.
In addition to the two of GR ($\Lambda, a_0$), the gqPG FLRW cosmology has 8 more physical parameters ($ a_2,\ a_3,\ \tilde \sigma_2,\ w_{3-2},\ w_{4+6},\ \mu_{3-2},\ w_{4-2}$).
 We used a manifestly homogeneous and isotropic Bianchi representation to develop an effective Lagrangian, an energy expression, and the 2nd order dynamical equations for the 3 gauge potentials. An equivalent first order system of 6 equations for the observables was obtained.  Certain inequalities were noted that identify a preferred physical range of parameters.  The late time dynamical behavior for generic solutions within the physical parameter range was discussed and illustrated using numerical evolution.
It turns out that the gqPG has effectively just one more free parameter ($w_{4-2}$) beyond what the earlier works~\cite{BHN,HN,AdP} considered; typical dynamical effects of this parameter were discussed and illustrated via numerical evolution.

The two dynamic connection modes do not directly couple to any known matter.
Hence they could be relatively large and unnoticed.
 The $0^+$ scalar mode couples directly to the Hubble expansion rate.
 Torsion could play a role in the observable universe.
 It might help in the understanding of some of the present puzzles associated with inflation, dark matter, and dark energy.  These topics should be explored within the general quadratic PG cosmology presented here.

\section*{Acknowledgements}

This work was supported in part by the Ministry of Science
and Technology under Grant No.\ MOST104-2112-M-006-020, and by the Headquarters
of University Advancement at the National Cheng Kung University,
which is sponsored by the Ministry of Education, Taiwan, ROC.

\end{document}